\documentclass[12pt]{iopart}
\usepackage{graphicx}
\usepackage[breaklinks=true,colorlinks=true]{hyperref}
\usepackage{cite}
\usepackage{xcolor}
\expandafter\let\csname equation*\endcsname\relax
\expandafter\let\csname endequation*\endcsname\relax
\usepackage{bm,latexsym,amssymb,amsmath,amsfonts}
\hypersetup{linkcolor=red, citecolor=blue, urlcolor=blue}

\def\Journal#1#2#3{\href{https://doi.org/#3}{{\it #1} #2}}
\def\arXiv#1#2{\href{https://arxiv.org/abs/#1}{[arXiv:#1 [#2]]}}

\begin{document}

\title[Imaging the destruction of a rotating regular black hole]{Imaging the destruction of a rotating regular\\black hole}

\author{M.~F.~Fauzi$^1$, H.~S.~Ramadhan$^{1,\star}$, A.~Sulaksono$^1$, Hasanuddin$^2$}
\address{$^1$Departemen Fisika, FMIPA, Universitas Indonesia, Depok 16424, Indonesia}
\address{$^2$Program Studi Fisika, FMIPA, Universitas Tanjungpura, Pontianak 78124, Indonesia}

\eads{\mailto{muhammad.fahmi31@ui.ac.id}, \mailto{hramad@sci.ui.ac.id}, \mailto{anto.sulaksono@sci.ui.ac.id}, \mailto{hasanuddin@physics.untan.ac.id}}

\vspace{10pt}
\begin{indented}
\item[]$\star$ Corresponding author
\end{indented}

\begin{abstract}
A regular black hole, unconstrained by the weak cosmic censorship conjecture, can exceed its critical spin limit and transition into a superspinar. In this paper, we investigate the observational appearance of a rotating regular black hole, specifically the Ghosh black hole and its superspinar counterpart, when surrounded by a thin accretion disk. The resulting images reveal distinct features: the black hole closely resembles its Kerr counterpart with slight deviations, while the superspinar configuration exhibits an inner photon ring structure. Furthermore, we investigate the image transition of the Ghosh black hole that has recently been destroyed by a collapsing null shell carrying a specific angular momentum. The results indicate that, apart from a possible sudden burst of light, the inner photon ring undergoes gradual transitions over time, with the transition times depending on the additional angular momentum gained by the black hole. Our findings also suggest that the transition timescale becomes significant for supermassive black holes, with masses at least less than about twice that of M87*.
\end{abstract}

\vspace{2pc}

\section{Introduction}

Black holes (BHs), as predicted by general relativity (GR), offer profound insights into the nature of spacetime and gravity. Observationally, the study of BHs has entered a groundbreaking era with the advent of the Event Horizon Telescope (EHT). The EHT has successfully captured images of supermassive BHs such as M87*~\cite{EventHorizonTelescope:2019dse,EventHorizonTelescope:2024dhe} and Sgr A*~\cite{EventHorizonTelescope:2022wkp}. These images reveal a central shadow surrounded by a bright, donut-shaped emission structure, consistent with the presence of an event horizon and an encircling accretion disk. To further test GR, numerous studies have simulated BH shadows and images~\cite{Lamy:2018zvj, Held:2019xde, Li:2024ctu, Yang:2024nin, He:2025rjq,Bambi:2019tjh,Vagnozzi:2022moj,Ghosh:2022gka,Igata:2025glk,Hou:2022eev,Zhang:2024lsf}, exploring potential deviations from the Kerr (or Kerr-Newman) solutions and providing valuable predictions for future observations. 

Theoretically, however, BHs still present fundamental challenges. Their solutions generically exhibit two types of `infinities': the event horizon(s), associated with infinite time dilation, and the spacetime singularity, where curvature diverges. The existence of horizons is constrained by the {\it weak cosmic-censorship conjecture} (WCCC), which asserts that spacetime singularities must remain hidden behind an event horizon~\cite{Penrose:1969pc}. For instance, the Kerr-Newman solution is subject to the condition $M^2>a^2 + Q^2$, where $M$, $a$, and $Q$ represent the BH’s mass, angular momentum (spin), and electric charge, respectively. If this bound is violated, the horizon disappears, exposing a {\it naked singularity}\textemdash a scenario that challenges the very foundation of classical GR.

One approach to addressing the singularity problem is the introduction of {\it regular black holes} (RBHs), which have primarily been studied in static cases. The first non-singular BH solution was proposed by Bardeen~\cite{Bardeen}, who introduced a new parameter in the metric to regulate the singularity by replacing it with a regular de Sitter core. Later, Ayon-Beato and García demonstrated that the Bardeen metric could be interpreted as a charged BH sourced by {\it non-linear electrodynamics} (NLED)~\cite{Ayon-Beato:2000mjt}.  Various mechanisms have since been proposed to construct RBH solutions, including NLED~\cite{Ayon-Beato:2000mjt, Ayon-Beato:1998hmi, Bronnikov:2000vy, Burinskii:2002pz, Bronnikov:2021uta}, de Sitter vacuum modifications~\cite{Hayward:2005gi,Cadoni:2022vsn}, and quantum gravity effects~\cite{Nicolini:2023hub,Ashtekar:2023cod,Eichhorn:2022bgu}. A distinctive feature of RBHs is that they are not constrained by the WCCC, allowing for horizonless configurations~\cite{Carballo-Rubio:2025fnc}. Once a static regular metric is obtained, its rotating counterpart can be derived using the Newman-Janis algorithm~\cite{Newman:1965tw}. 

A class of static NLED-based RBH solutions has been proposed in various studies by modifying the Schwarzschild metric through an exponential-type probability distribution~\cite{Balart:2014cga, Neves:2014aba, Azreg-Ainou:2014pra,Culetu:2014lca}. Using the Newman-Janis transformation, Ghosh later derived its rotating counterpart~\cite{Ghosh:2014pba}. This rotating solution exhibits significant deviations from the classical Kerr-Newman solution, particularly in its horizon structure, ergoregion, and photon sphere. For instance, the photon ring structure differs significantly from that of Kerr BHs, highlighting distinct characteristics in the photon sphere's properties~\cite{Kumar:2020ltt}. Further analysis of the ergosphere and shadow reveals that their sizes and shapes are highly sensitive to parameters such as charge and spin, resulting in a richer structure compared to the Kerr solution~\cite{Ghosh:2020ece}. Additionally, studies of spherical orbits (both photon and timelike) indicate that the corresponding trajectories exhibit larger latitudinal oscillation amplitudes as the Ghosh parameter increases, underscoring the influence of this parameter on orbital dynamics~\cite{Alam:2024mmw}.

Another particularly intriguing study was done by Eichhorn and Held~\cite{Eichhorn:2022bbn}, where they examined RBHs within the framework of asymptotic safety gravity~\cite{Eichhorn:2022bgu, Platania:2023srt}. Their findings suggest that {\it overspinning}, a scenario in which a BH's angular momentum exceeds its critical value (leading to the destruction of the event horizon), could have profound observational consequences. Specifically, the resulting image exhibits additional ring-like secondary images within the primary shadow, effectively lighting up the BH and increasing the overall observed intensity. This brightening occurs instantaneously, as the horizon is globally resolved in the modified spacetime. Furthermore, instead of a gradual transition, the intensity undergoes a sudden, discontinuous shift once the angular momentum surpasses the critical threshold.

In this study, we adopt the definition of \textit{superspinar} as proposed in Ref.~\cite{Torres:2024eli}: rotating BH spacetime with angular momentum exceeding a critical threshold, $a^2>a_c^2$. In the classical Kerr solution, such a configuration corresponds to a naked singularity, which may be rendered physically viable through string-theoretic effects~\cite{Gimon:2007ur, Nguyen:2023clb, Stuchlik:2010zz}. In the case of rotating RBH, the superspinar solution might imply something entirely different, depending on the theory used to construct the metric. In both cases, the superspinar could potentially be obtained by \textit{destroying} the BH, {\it i.e.,} resolving the horizon by sending a test particle with specific angular momentum to be captured by the extremal or near-extremal BH~\cite{Saa:2011wq,Siahaan:2015ljs, Siahaan:2022fht,Li:2013sea}. However, within GR itself, the destruction of rotating BHs is generally not allowed in most scenarios studied in the literature~\cite{Wald:1974hkz,Barausse:2010ka,Sorce:2017dst}.

In this work we focus on analyzing the appearance of Ghosh BHs and their superspinar counterparts when surrounded by a thin accretion disk. Inspired by the lighting-up scenario~\cite{Eichhorn:2022bbn}, we take a different approach to image transition; we simulate the image transition over a time interval following the recent destruction of a RBH by a collapsing null shell with a certain angular momentum, which simplifies the calculations. Furthermore, we investigate whether the image intensity transition time is relevant to our observations.

This paper is organized as follows. In Sec.~\ref{sec. regular rotating black hole}, we provide a brief overview of the rotating RBH model employed in our study, along with the proposed destruction scenario. In Sec.~\ref{sec. geodesics around axially}, we introduce the Hamiltonian formalism used for numerically computing null geodesics, which serves as the foundation for our ray-tracing procedure to generate BH images. In Sec.~\ref{sec. optical appearance}, we briefly discuss the accretion disk model, and proceed to the image results and analysis. Sec.~\ref{sec. image transition} is devoted to examining the image transition scenario, presenting the corresponding simulated images, and assessing their observational implications. Finally, in Sec.~\ref{sec. conclusion}, we summarize our findings and provide concluding remarks.

\section{Regular rotating black hole and superspinar}
\label{sec. regular rotating black hole}

\subsection{The regular black hole spacetime}

A general form of a rotating BH metric in Boyer-Lindquist coordinates is given by
\begin{align}
ds^2 =&-\left(1-\frac{2m(r)r}{\Sigma}\right)dt^2 - \frac{4am(r)r}{\Sigma}\sin^2 \theta dtd\phi\nonumber\\ &+\Sigma\left(\frac{dr^2}{\Xi} + d\theta^2\right) + \frac{\mathcal{A}}{\Sigma}\sin^2 \theta d\phi^2,
\label{eq. metric axisymmetric}
\end{align}
with
\begin{align}
	\Sigma\equiv&r^2 + a^2 \cos^2 \theta,\nonumber\\
	\Xi\equiv& r^2 + a^2 - 2m(r)r,\nonumber\\
	\mathcal{A}\equiv&\left(r^2 + a^2\right)^2-\Xi a^2 \sin^2 \theta,
\end{align}
and $a$ is the spin parameter. Ghosh constructed a rotating RBH by adopting a mass function of the form~\cite{Ghosh:2014pba,Culetu:2014lca}
\begin{equation}
    m(r)=Me^{-k/r},
    \label{eq. mass ghosh}
\end{equation}
where $M$ is the mass and $k$ is the new parameter introduced to regularize the metric. Notably, all curvature invariants remain finite everywhere. In particular, the Kretschmann scalar, 
\begin{equation}
    K\equiv R^{\alpha\beta\gamma\kappa}R_{\alpha\beta\gamma\kappa}\rightarrow0,
\end{equation}
as $r\rightarrow0$, ensuring the absence of a central singularity. 

\begin{figure}[htbp!]
    \centering
\includegraphics[width=0.7\linewidth]{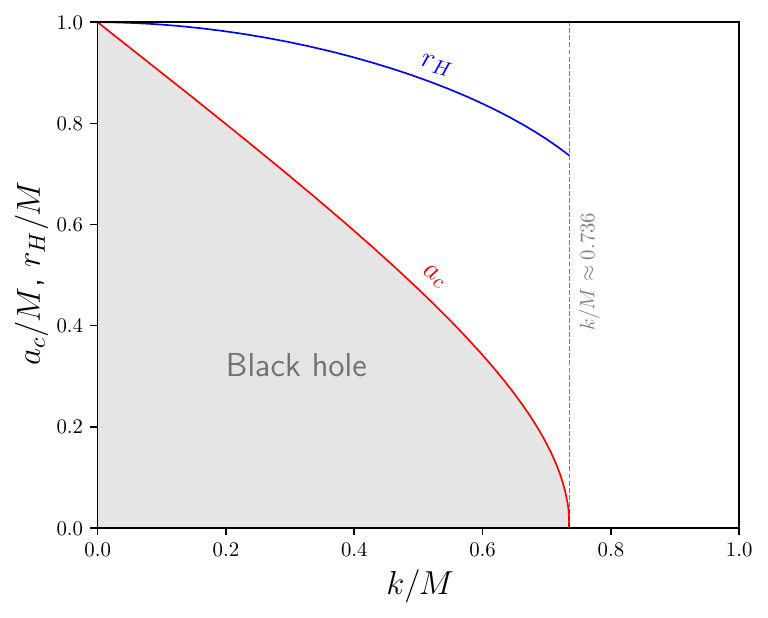}
    \caption{The critical spin value $a_c/M$ (red line) and its corresponding extremal horizon radius $r_H^e/M$ (blue line) as functions of $k/M$. The grey shaded area depicts the BH configuration where the spacetime contains horizon(s), while the region outside (with $a>0$) corresponds to the superspinar configuration.}
    \label{fig. extremum value k a}
\end{figure}

The event horizon radii $r_H^\pm$ are determined by the condition $\Xi=0$, which leads to the equation
\begin{equation}
    r_H^\pm+a^2-2Me^{-k/r_H^\pm}=0.
\end{equation}
These roots can be obtained numerically. In general, similar to its singular counterpart, an RBH can exhibit three possible configurations: two horizon BH ($a<a_c$), extremal BH ($a=a_c$), and horizonless ($a>a_c$)\textemdash which, in our case, is dubbed as the superspinar. The extremal condition is obtained by simultaneously solving 
\begin{equation}
	\left.\Xi\right|_{r=r_H^e} = \left.\partial_r \Xi\right|_{r=r_H^e} = 0,  
\end{equation} 
where $r_H^e$ is the extremal horizon radii. For the Ghosh mass function, there exists an upper bound on the parameter $k$ required to support a BH configuration. In the non-rotating limit ($a=0$), this threshold is found to be approximately $k\approx0.736M$. The extremal values of the configuration in the parameter space of $k$ and $a$ are illustrated in Fig.~\ref{fig. extremum value k a}.

\subsection{Spinning up the black hole beyond extremality}

For the sake of simplicity, let us consider a collapsing null shell with a particular angular momentum, which is then ``transferred" to the BH. To describe this, we incorporate the advanced/ingoing null coordinate, which can be obtained by applying the following transformations to Eq.~\eqref{eq. metric axisymmetric}:
\begin{equation}
    u\equiv t+\int\frac{r^2+a^2}{\Xi}dr,\qquad \varphi\equiv \phi+\int \frac{a}{\Xi}dr. \label{eq. coord null transform}
\end{equation}
This coordinates allow the spacetime to depend on $u$, typically in the mass function such that $m(r)\to m(r,u)$. In our case, since the collapsing null shell changes the BH's final angular momentum, we also allow the spin parameter to depend on $u$, \textit{i.e.}, $a \to a(u)$. Thus, the line element in this coordinate becomes
\begin{align}
    ds^2=&-\left(1-\frac{2m(r,u)r}{\Sigma}\right)du^2+2dudr+\Sigma d\theta^2 - 2a(u)\sin^2\theta drd\varphi \notag\\ 
    &- \frac{4a(u)m(r,u)r}{\Sigma}\sin^2\theta dud\varphi+\frac{\mathcal{A}}{\Sigma}\sin^2\theta d\varphi^2, \label{eq. null coord}
\end{align}
which has the same form as the horizon penetrating coordinates used in Ref.~\cite{Eichhorn:2022bbn}.

Again, for simplicity, we assume that the collapsing null shell at $u = u_c$ changes the BH spin parameter from $a_0 \to a_0 + \delta$ \textit{almost} instantaneously, while keeping the mass profile unchanged, so that $m(r,u) \to m(r)$. We then demand that $a(u)$ be roughly defined as
\begin{equation}
    a(u)\approx\begin{cases}
        a_0,&u<u_c\\
        a_0+\delta.&u\geq u_c
    \end{cases}
\end{equation}
However, the derivative of this step function with respect to $u$ at $u=u_c$ yields a Dirac delta function, which may lead to a singularity spikes in the curvature scalar at $\sim u_c$ and not be properly handled by our numerical procedure to compute the null geodesics. Therefore, we instead take the form of $a(u)$ to be
\begin{equation}
    a(u)=a_0 + \delta \Theta(u-u_c),\qquad \Theta(u-u_c)=\left[1+\exp\left(-\frac{u-u_c}{\sigma_s}\right)\right]^{-1},
    \label{eq. spin func}
\end{equation}
where the $\Theta(u-u_c)$ is known as the sigmoid function, and $\sigma_s$ determines the smoothness of the transition from $a_0\to a_0+\delta$, interpreted as the inverse of the \textit{spin transition rate}. Instantaneous transition implies $\sigma_s\to0$.

\begin{figure}[htbp!]
    \centering
\includegraphics[width=0.8\linewidth]{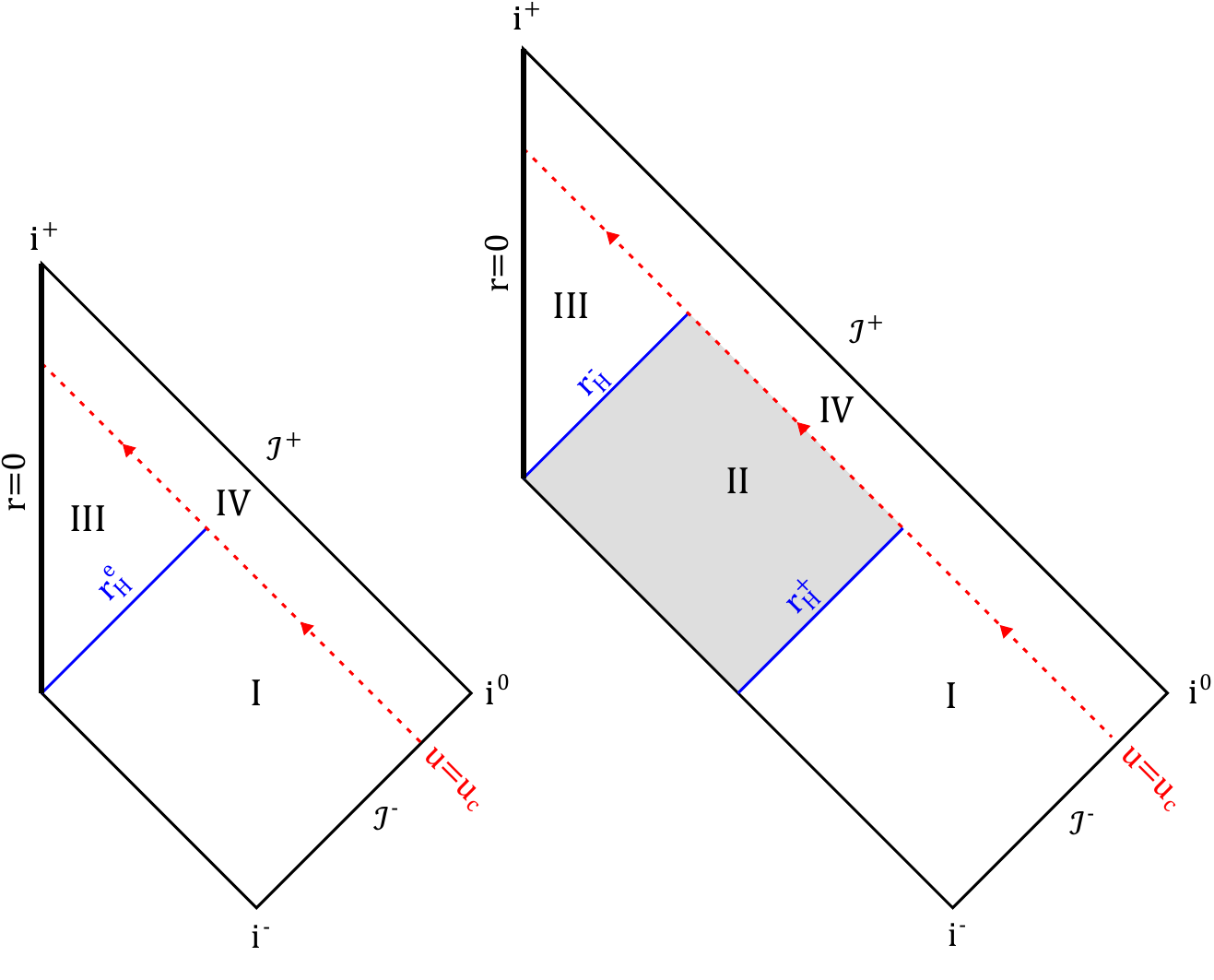}
    \caption{A sketch of the Penrose diagram for extremal (left) and non-extremal (right) BH destruction with a collapsing null shell (red dashed lines). Region I represents the spacetime exterior to the BH, while Region II is the region between the outer horizon $r_H^+$ and the inner horizon $r_H^-$. In the extremal BH, $r_H^-$ and $r_H^+$ coincide at $r_H^e$, and thus Region II does not exist. Region III corresponds to the BH interior, and Region IV denotes the superspinar spacetime.}
    \label{fig. penrose diagram destroying BH}
\end{figure}

A spin addition to the BH that satisfies $a_0 + \delta > a_c$ would lead to the absence of solutions for $\Xi = 0$, causing the horizon to disappear and thereby destroying the BH. The entire horizon area dissolves globally \textit{immediately} after the spin parameter exceeds its critical limit, even if the spin parameter increases gradually. For the instantaneous transition at $u = u_c$, we present a sketch of the Penrose diagram for both extremal and non-extremal BH destruction in Fig.~\ref{fig. penrose diagram destroying BH}, based on those in Ref.~\cite{Torres:2022twv}.

\begin{figure}[htbp!]
    \centering
\includegraphics[width=0.95\linewidth]{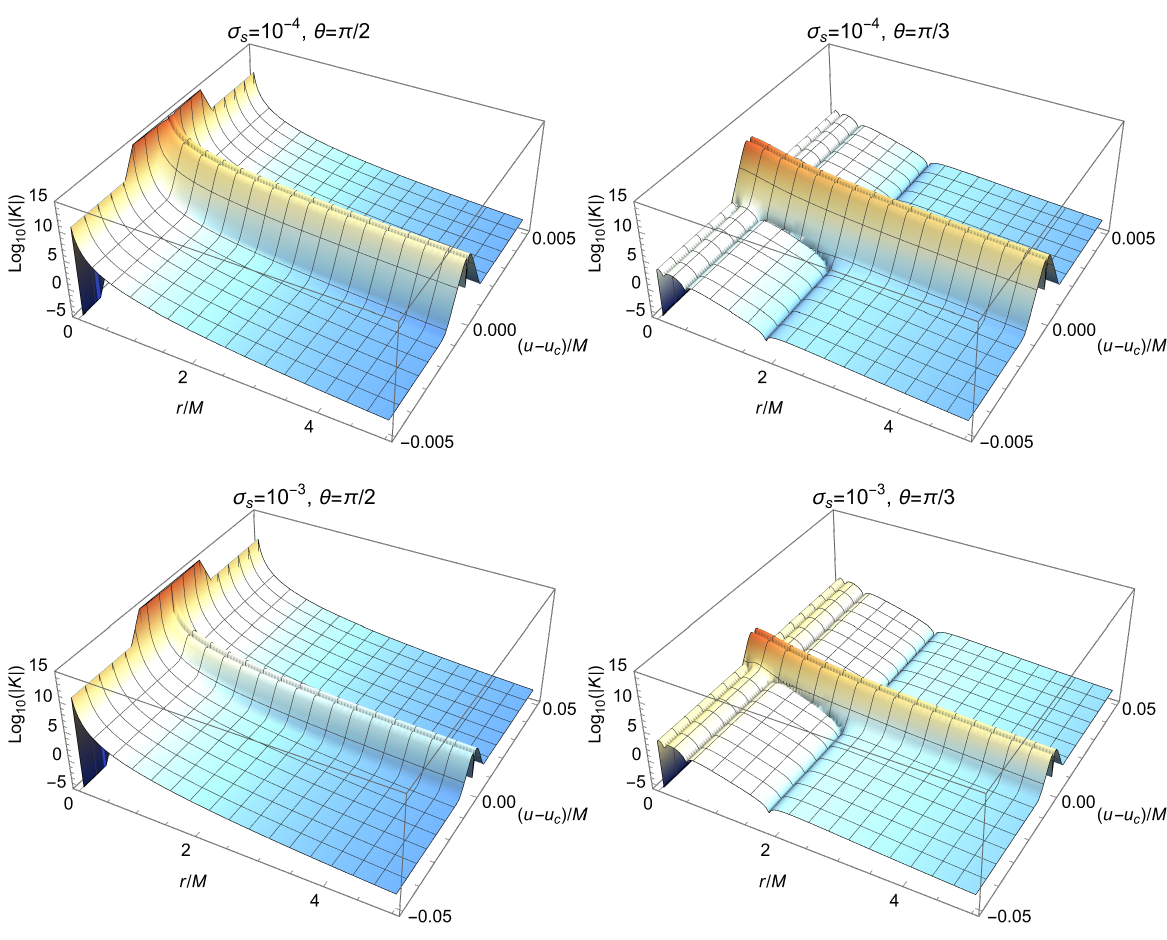}
    \caption{The absolute value of the Kretschmann scalar in logarithmic scale for $\delta/M = 5 \times 10^{-3}$, $k/M=0.01$, and $a_0=a_c$, with $\sigma_s = 10^{-4}$ (left column) and $\sigma_s = 10^{-3}$ (right column), shown as a function of $r$ and $u$ near $u_c$.}
    \label{fig. kretschmann dyn}
\end{figure}

\begin{figure}[htbp!]
    \centering
\includegraphics[width=1\linewidth]{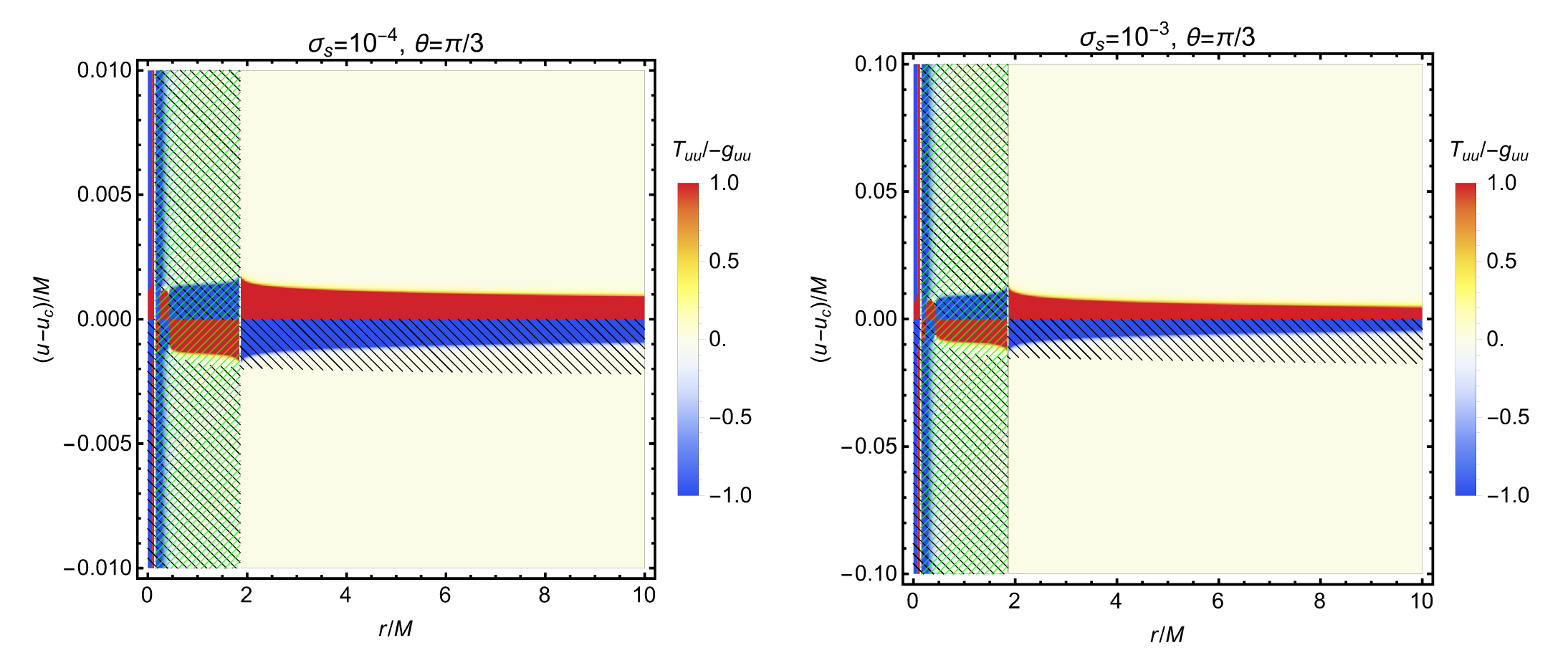}
    \caption{The quantity $T_{uu}/(-g_{uu})$ at $\theta = \pi/3$ for $\delta/M = 5 \times 10^{-3}$, $k/M = 0.01$, and $a_0 = a_c$, with $\sigma_s = 10^{-4}$ (left column) and $\sigma_s = 10^{-3}$ (right column), plotted as a function of $r$ and $u$ near $u_c$. The black dashed region indicates where the WEC is violated, while the green dashed region marks the ergoregion where $-g_{uu} \leq 0$.}
    \label{fig. WEC dynamic}
\end{figure}

Additionally, we investigate the spacetime by studying its curvature scalar, specifically the Kretschmann scalar $K$, near the BH destruction event. The full expression of this quantity in the spacetime of Eq.~\eqref{eq. null coord} with the mass function of Eq.~\eqref{eq. mass ghosh} is rather cumbersome, and we provide it in~\ref{appendix. kretschmann}. The scalar contains the first- and second-order derivatives of $a(u)$ with respect to $u$\textemdash which, for $a(u)$ given in Eq.~\eqref{eq. spin func}, are proportional to $\sigma_s^{-1}$ and $\sigma_s^{-2}$, respectively\textemdash and these terms are most prominent near $u = u_c$. This indicates the presence of a curvature spike at $u \sim u_c$, especially for small $\sigma_s$. Moreover, the spikes are further amplified by the quartic dependence on $a'(u)$ and quadratic dependence on $a''(u)$, making the curvature scalar proportional to $\sigma_s^{-4}$. As shown in Fig.~\ref{fig. kretschmann dyn}, for $\delta/M = 5 \times 10^{-3}$ the curvature spike with $\sigma_s \sim 10^{-4}$ reaches an order of $\approx 10^{15}$ near the BH core, occurring close to the null shell. In the case of an instantaneous transition with $\sigma_s = 0$, this would generate a singularity along the collapsing null shell at $u = u_c$. 

We also expect that the transition would lead to a violation of the weak energy condition (WEC) near the null shell. To satisfy the WEC, the inequality~\cite{Hawking:1973uf}
\begin{equation}
T_{\mu\nu}n^\mu n^\nu \geq 0,
\end{equation}
must hold, where $T_{\mu\nu}$ is the energy-momentum tensor and $n^\mu$ is any timelike vector satisfying $n_\mu n^\mu = -1$. For a straightforward calculation, we employ a stationary timelike vector ($n^r = n^{\theta} = n^{\varphi} = 0$), and the WEC is explicitly violated when $T_{uu}/(-g_{uu}) < 0$. This expression is valid only outside the ergoregion, \textit{i.e.}, for $-g_{uu} > 0$, since such a vector is not well defined when $-g_{uu} \leq 0$ (see, e.g., Ref.~\cite{Dorau:2024zyi}). The energy-momentum tensor corresponding to the metric in Eq.~\eqref{eq. null coord} is obtained by solving the field equations $G_{\mu\nu} = 8\pi T_{\mu\nu}$, and the full expression for the $T_{uu}$ component is provided in~\ref{appendix. kretschmann}. As shown in Fig.~\ref{fig. WEC dynamic}, $T_{uu}/(-g_{uu})$ takes negative values near $u \lesssim u_c$, and hence the collapsing null shell indeed violates the WEC.

\section{Geodesics around axially symmetric spacetime}
\label{sec. geodesics around axially}
 
The dynamics of photon around this RBH is determined by solving the corresponding null geodesic equations. Here, we adopt the Hamiltonian formalism to derive the equations of motion (EoM) and numerically integrate them, following the approach outlined in Refs.~\cite{Rezzolla:2013dea, Bacchini:2018zom}. Below, we provide a concise overview of the formalism and its implementation for our specific case.

In general, the rotating metric can be expressed in the so called $3+1$ decomposition form
\begin{equation}
    g_{\mu\nu} =
    \begin{pmatrix}
    -\alpha^2 + \beta_{k}\beta^{k} & \beta_i\\
    \beta_j & \gamma_{ij}
    \end{pmatrix},\label{eq. gmunu split}
\end{equation}
where $\alpha$ is the lapse function, $\beta_j$ is the shift three vector, and $\gamma_{ij}$ represents the spatial components of $g_{\mu\nu}$. Greek and Latin indices label the spacetime and spatial coordinates, respectively. The inverse metric take the form
\begin{equation}
    g^{\mu\nu} =
    \begin{pmatrix}
        -1/\alpha^2 & \beta^i/\alpha^2\\
        \beta^j/\alpha^2 & \gamma^{ij} - \beta^i\beta^j/\alpha^2
    \end{pmatrix},
\end{equation}
where $\gamma^{ij}$ is the algebraic inverse of $\gamma_{ij}$, and $\beta^j=\gamma^{ij}\beta_i$. 

This study incorporates the Hamiltonian formalism to obtain the geodesic equations. The general Hamiltonian is given by~\cite{Bakun:2024dwq}
\begin{equation}
    H=\frac{1}{2}g^{\mu\nu}p_\mu p_\nu=-\frac{1}{2}\epsilon,
    \label{eq. hamiltonian}
\end{equation}
where $p^\mu\equiv dx^\mu/d\tau=g^{\nu\mu}p_\nu$ is the four-velocity of the particle, and $\epsilon=0$ or $1$ correseponds to a null or massive test particle, respectively. The equations of motion (EoM) are expressed as
\begin{equation}
    \frac{dx^\mu}{d\tau}=\frac{\partial H}{\partial p_\mu},\qquad \frac{dp_\mu}{d\tau}=-\frac{\partial H}{\partial x^\mu}, \label{eq. dH x standard}
\end{equation}
where $\tau$ is an affine parameter. With the $3+1$ decomposition, one can obtain
\begin{equation}
    \frac{dx^0}{d\tau}=p^0 = \left(\frac{\gamma^{jk}p_jp_k+\epsilon}{\alpha^2}\right)^{1/2}. \label{eq. p0}
\end{equation}
It should be noted, however, that this decomposition might not be useful in some cases. For example, with a Vaidya spacetime, one finds $g_{rr}=0$ which leads to vanishing determinant of $\gamma_{ij}$, causing the lapse function $\alpha$ to diverge~\cite{Ma:2024kbu}. We also observe another issue with the lapse function $\alpha$: in the horizon-penetrating coordinate, our spacetime yields a negative $\alpha^2$, resulting in an imaginary $p^u$. Therefore, throughout this study, Eq.~\eqref{eq. p0} will be considered valid only for stationary spacetimes where $p^0=p^t$.

In the superspinar case, a particle may eventually reach the core at $r=0$. However, the geodesic equations in Boyer–Lindquist coordinates can yield solutions that extend into the negative radial region, \textit{i.e.} $r<0$. Fortunately, in RBH spacetimes, since the core is regular and does not suffer from differentiability issues, particles moving across the $r=0$ ``disk" remain within the $r>0$ portion of the spacetime~\cite{Torres:2022twv}. Therefore, considering that our spacetime is flat near the core region, geodesics that cross into the region with negative $r$ can be equivalently transformed into trajectories with $r\to r'=|r|$ and $\theta\to\theta'=\pi-\theta$, allowing the particle motion to be smoothly continued across the ``disk."

The geodesic equations are solved numerically using a standard Runge–Kutta–Fehlberg (RKF45) method with adaptive stepping. The relative error of all integration components is maintained at the level of $\sim 10^{-7}\text{–}10^{-5}$ by adjusting the integration step size. In the BH destruction scenario, there are two crucial regions where we decrease the error tolerance by an order of $10^{-5}$ relative to the initial tolerance: (i) near the (destroyed) horizon, defined by the points where $|\Xi/\Sigma| < 10^{-1}$, and (ii) near the destruction moment at $|u - u_c|/M < 5$. This ensures that the null geodesics are obtained more efficiently while preserving relatively high accuracy.

\subsection{Stationary spacetime}

The Hamiltonian formalism can be further simplified in stationary spacetimes. The resulting simplified Hamiltonian for a test particle is given in~\cite{Gourgoulhon:2012ffd}
\begin{equation}
    \tilde{H} = \alpha\left(\gamma^{jk}p_jp_k + \epsilon\right)^{1/2} - \beta^j p_j.
    \label{eq. hamiltonian modified}
\end{equation}
The corresponding $x^0$-dependent EoM are given by
\begin{equation}
    \frac{dx^i}{dt} = \frac{\partial \tilde{H}}{\partial p_i}, \qquad
\frac{dp_i}{dt}=-\frac{\partial \tilde{H}}{\partial x^i}.\label{eq. dH x}
\end{equation}
Eq.~\eqref{eq. dH x} govern the geodesic motion of (timelike and null) particles in stationary spacetimes. It is also known that $|\tilde{H}|=|p_0|$~\cite{Bacchini:2018zom}, with $p_0$ is the covariant form of $p^0$ ($p_0=g_{\mu0}p^\mu$).

One key advantage of this simplified formalism is its ability to directly compute the trajectory in terms of coordinate time $t$ numerically. Instead of solving four coupled second-order ordinary differential equations, often rewritten as eight coupled first-order equations in numerical computations, this approach reduces the system to just six first-order equations for the spatial components $x^i=(r,\theta,\phi)$, while the coordinate time $t$ is implicitly accounted for through the integration step size.

\subsection{Dynamical spacetime}

The simplified Hamiltonian formalism given in Eqs.~\eqref{eq. hamiltonian modified} and~\eqref{eq. dH x} is no longer valid in dynamical spacetime. We adopt the standard formalism from Eqs.~\eqref{eq. hamiltonian} and~\eqref{eq. dH x standard}, solving numerically for all four variables $x^{\mu}=(u,r,\theta,\varphi)$.

The initial velocity is assigned using Eq.~\eqref{eq. coord null transform}
\begin{equation}
    p^u_{(0)}=p^t_{(0)}+\left(\frac{a(u_{(0)})^2+r_{(0)}^2}{\Xi_{(0)}}\right)p^r_{(0)},\qquad p^{\varphi}_{(0)}=p^\phi_{(0)} + \frac{a(u_{(0)})}{\Xi_{(0)}}p^r_{(0)},
\end{equation}
and Eq.~\eqref{eq. p0} is used to determine $p^t$. The subscript $(0)$ denotes the initial condition at the observer. Assuming a flat spacetime at the distant observer, we apply the approximation
\begin{equation}
    \lim_{r\to r_{(0)}}\frac{r^2+a^2}{\Xi}\approx1,\qquad \lim_{r\to r_{(0)}}\frac{a}{\Xi}\approx0,
\end{equation}
so that we can set $u_{(0)}=t_{(0)}+r_{(0)}$ and $\varphi_{(0)}=\phi_{(0)}$ as the initial conditions at the observer, where the $t_{(0)}$ determines the coordinate time at the observer. We have confirmed that the images obtained using this general Hamiltonian method coincide with those obtained from the simplified method, with the only difference being a slightly longer computation time.

\section{Optical appearance involving thin accretion disk}
\label{sec. optical appearance}

\subsection{Thin accretion disk}

Simulating BH images generally requires a ray-tracing procedure, which involves computing photon trajectories or null geodesics. In this study, we assume an optically and geometrically thin accretion disk with monochromatic emission. An optically thin accretion disk is transparent, allowing light to pass through without significant absorption or scattering, a characteristic often associated with supermassive BHs~\cite{Gold:2020iql}. The geometrically thin assumption confines the disk to an infinitesimally narrow region in the equatorial plane ($\theta = \pi/2$), significantly simplifying the detection algorithm. Furthermore, we incorporate the effects of accretion flow and gravitational redshift, which will be discussed in detail later.

\subsubsection{Intensity profile}

The accumulation of matter in an accretion disk typically results in electromagnetic emission. For simplicity, we consider an emission profile $I_e(r)$, which depends on the matter density and temperature of the disk. We adopt the Gralla-Lupsasca-Marrone (GLM) model \cite{Gralla:2020srx}, given by
\begin{equation}
	I_e(r) = \frac{\exp\left\{-\frac{1}{2}\left[\gamma + \operatorname{arcsinh}{\left(\frac{r-\mu}{\sigma}\right)}\right]^2\right\}}{\sqrt{\left(r-\mu\right)^2+\sigma^2}},
	\label{eq. intensity profile glm}
\end{equation}
where $\gamma$, $\mu$, and $\sigma$ are parameters that determine the shape of the emission profile. Compared to exponential cut-off models (see, e.g., Refs.~\cite{Meng:2023htc,Zeng:2023fqy,Guo:2022iiy}), the GLM profile offers a continuous intensity function with tunable smoothness, controlled by the parameters $\sigma$ and $\gamma$. This model was found to closely match the observational predictions for the intensity profiles of astrophysical accretion disks, as obtained through general relativistic magnetohydrodynamics simulations~\cite{Vincent:2022fwj}, and has been used in several other investigations of BHs and ultracompact objects' appearance.~\cite{Rosa:2023hfm, Rosa:2023qcv, Rosa:2024bqv, daSilva:2023jxa,Macedo:2024qky,Fauzi:2024nta}. We set these parameters as $\gamma = -2$, $\mu = R_{ISCO}$, and $\sigma = M/4$, where $R_{ISCO}$ denotes the radius of the {\it innermost stable circular orbit} (ISCO) for a massive particle. This model assumes that emission peaks near the ISCO radius and ceases beyond it, as any massive particle crossing within the ISCO inevitably spirals toward the BH, leading to a peak in matter density around this radius.

Additionally, we provide a brief discussion and comparison with the disk emission models used in other studies. Various terms are used in the literature to describe the emission profile of accretion disks, but here we focus specifically on the emissivity function. In Ref.~\cite{Hou:2022eev}, a logarithmic-exponential emissivity profile is employed, which has been used to fit simulation data of the 230 GHz images of M87*~\cite{Chael:2021rjo}. Furthermore, Refs.~\cite{Hou:2023bep,Zhang:2024lsf} presents an approximate analytical emissivity derived from magnetohydrodynamic simulations, expressed in a more complex form that includes the electromagnetic field strength on the disk. They also considered a thick accretion disk, in which the emissivity depends on the polar coordinate $\theta$.  However, despite the simplicity of the GLM model, we consider it particularly useful for studying the image features of a BH surrounded by an accretion disk.

In Refs.~\cite{Hou:2022eev,Hou:2023bep,Zhang:2024lsf}, the accretion disk profile extends beyond the ISCO radius in an exponential fashion down to the event horizon, requiring the plunging region to be taken into account when modeling the accretion flow within the ISCO. This approach contrasts with the GLM model used in our study, where the accretion disk terminates near the ISCO, allowing us to neglect the plunging region entirely.

The accretion disk intensity profile used in this study comes with certain limitations, particularly in the superspinar configuration. While the GLM model may successfully describe the disk intensity in a rotating BH spacetime, this may not hold true in the superspinar case due to the absence of an event horizon. In horizonless spacetimes, matter may pass through the central region and accumulate within, potentially generating additional disk intensity that could be superimposed on the emission from circularly orbiting material~\cite{Rosa:2024bqv,Rosa:2023hfm,Rosa:2023qcv}. Despite this possibility, we adopt the GLM model in our phenomenological analysis for the sake of simplicity, as a more complete treatment lies beyond the scope of this paper.

\subsubsection{Accretion flow and innermost stable circular orbit}

We assume the accretion flow to be Keplerian on the equatorial plane, described by the (timelike) 4-velocity $v^{\nu}_e$
\begin{equation}
    v^{\nu}_e = (v^t_e, v^r_e, 0, v^\phi_e). 
\end{equation}
Additionally, we assume that the matter in the disk does not experience any radial drift, implying $v^r_e = 0$. This assumption also implies that the plunging region of the accretion disk is not considered in our analysis. The azimuthal component $v^\phi_e$ is given by~\cite{Ryan:1995wh,Ozel:2021ayr}
\begin{align}
v^\phi_e=&\frac{\Omega}{\sqrt{-g_{tt}-(2g_{t\phi}+g_{\phi\phi}\Omega)\Omega}},\\
\Omega=&\frac{-g_{t\phi,r} \pm \sqrt{g_{t\phi,r}^2 - g_{tt,r}g_{\phi\phi,r}}}{g_{\phi\phi,r}},
\end{align}
where the $\pm$ sign corresponds to the direction of the accretion flow orbit: prograde ($+$) or retrograde ($-$). We restrict our analysis to prograde accretion flow, where the flow direction aligns with the BH's spin. Imposing the 4-velocity condition,
\begin{equation}
    g_{tt} (v^t_e)^2 + g_{\phi\phi}(v^\phi_e)^2 + 2g_{t\phi}v^t_e v^\phi_e = -1,
\end{equation}
we obtain the time component $u^t$ as
\begin{equation}
    v^t_e = \frac{g_{t\phi} v^\phi \pm \sqrt{-g_{tt} + (g_{t\phi}^2 - g_{tt} g_{\phi\phi})(v^{\phi}_e)^2}}{-g_{tt}},
    \label{eq. ut accretion disk}
\end{equation}
where we select only the positive root.

The 4-velocity components of the accretion flow determine the redshift effects,  denoted by the factor $\tilde{g}$, which include both the gravitational redshift and the relativistic Doppler effect. A source emitting light at frequency $\nu_e$ will experience a redshift as it propagates toward the observer, resulting in an observed frequency $\nu_o$. Their relationship is given by~\cite{Ryan:1995wh,Ozel:2021ayr,M:2022pex}
\begin{equation}
    \nu_o = \tilde{g} \nu_e,\qquad \tilde{g} = \frac{-k_\mu v^\mu_o}{-k_\nu v^\nu_e},
\end{equation}
where the subscript $e$ and $o$ refer to the emitting frame (accretion disk) and observer frame, respectively, and $k_\mu = k^{\nu}g_{\mu\nu}$ is the photon 4-momentum (null vector). Since we assume a static, distant observer, the observer's 4-velocity is given by $v_o^\mu = (1,0,0,0)$. Using the 4-velocity of the accretion flow, the redshift factor simplifies to
\begin{equation}
    \tilde{g} = \left(v^t_e + \frac{k_\phi}{k_t}v^\phi_e\right)^{-1}.
\end{equation}
The observed intensity is determined using the Lorentz-invariant relation $I^\nu_e/\nu_e^3 = I^\nu_o/\nu_o^3$. 
Integrating over all frequencies gives the total observed intensity
\begin{equation}
    I_o(r) = \int I^\nu_o(r) d\nu_o = \tilde{g}^4 I_e(r).
\end{equation}

The accretion flow features an ISCO, whose radius is determined by solving for a point where $u^r = \partial_r V_t(r) = \partial_{rr} V_t(r) = 0$, where $V_t(r)$ represents the effective potential for timelike geodesics~\cite{Rumler:2024cmb},
\begin{equation}
    V_t(r) = -\frac{1}{2r^2}\left(g_{tt}L^2 + g_{t\phi}EL + g_{\phi\phi}E^2\right) + \frac{1}{2}
\end{equation}
The $L$ and $E$ are integration constants, and for circular orbits, they can be determined by
\begin{align}
    L =& \frac{g_{\phi\phi}\Omega + g_{t\phi}}{\sqrt{-g_{\phi\phi}\Omega^2 - 2g_{t\phi} \Omega - g_{tt}}},\nonumber\\
    E =& -\frac{g_{t\phi}\Omega + g_{tt}}{\sqrt{-g_{\phi\phi}\Omega^2 - 2g_{t\phi} \Omega - g_{tt}}}.\label{eq. E ISCO}
\end{align}
For stable circular orbits, we require
\begin{equation}
    \left.g_{tt,rr} L^2 + 2g_{t\phi,rr} LE + g_{\phi\phi,rr} E^2 - 2\right|_{r = R_{ISCO}} \leq 0
    \label{eq. determine ISCO}
\end{equation}
and this quantity is zero at the ISCO. Hence, substituting Eq.~\eqref{eq. E ISCO} into Eq.~\eqref{eq. determine ISCO} and solve for $r$, we will eventually obtain the $R_{ISCO}$.

\begin{figure}[htbp!]
    \centering
\includegraphics[width=0.5\linewidth]{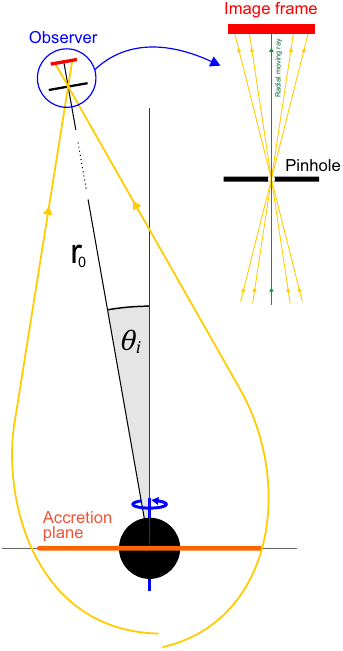}
    \caption{A 2D schematic of the pinhole ray-tracing procedure. Light rays (yellow curves) are traced backward from the observer, located at a distance $r_0$ from the BH, and directed toward the BH. The accretion disk (orange thick line) lies in the equatorial plane ($\theta=\pi/2$) and follows a retrograde orbit relative to the spin of the BH.}
    \label{fig. observer illustration}
\end{figure}

\subsection{Ray tracing and image results for stationary black hole}

To generate simulated images of curved spacetime, we employ a ray-tracing procedure that involves computing null geodesics using the Hamiltonian formalism discussed in Sec.~\ref{sec. geodesics around axially}, utilizing the backward-integration method. Specifically, we adopt the pinhole ray-tracing technique, where light rays are traced backward from an observer positioned at near-infinity, for instance, at $r_{(0)} = 500M$ and $\phi_{(0)}=0$. Each time a ray intersects the accretion disk on the equatorial plane, the corresponding intensity at that point\textemdash accounting for redshift effects\textemdash is summed and mapped to the appropriate pixel on the image frame $(x_p,y_p)$. Mathematically, the total intensity at each pixel point $I_s(x_p,y_p)$ is given by
\begin{equation}
    I_s(x_p,y_p)=\sum_{n_i=1}^{N_i}I^{(n_i)}_o(r),
\end{equation}
where $n_i=1,...N_i$ denotes the number of intersections of a light ray with the equatorial plane, and $I_o^{(n_i)}$ is the observed intensity at the $n_i$-th intersection. The ray-tracing scheme is illustrated in Fig.~\ref{fig. observer illustration}.

\begin{table}[htbp!]
    \centering
    \begin{tabular}{cclccccccc} \hline \hline
        Figure & \hspace{0.4cm} & \multicolumn{1}{c}{Spacetime} & \hspace{0.4cm} & \multicolumn{1}{c}{$a/a_c$} & \hspace{0.4cm} & \multicolumn{1}{c}{$k/M$} & \hspace{0.4cm} & \multicolumn{1}{c}{$r_{ISCO}/M$} & \hspace{0.4cm}\\ \hline
        Fig.~\ref{fig. 03rad redshifted images} & & Ghosh & & $0.90$ & & $0.01$ & & $2.32$ & \hspace{0.4cm}\\
         & & & & $0.99$ & &  & & $1.45$ & \hspace{0.4cm}\\
         & & & & $1.01$ & &  & & $0.75$ & \hspace{0.4cm}\\
         & & & & $1.20$ & &  & & $0.69$ & \hspace{0.4cm}\\
         & & & & $0.90$ & & $0.20$ & & $2.20$ & \hspace{0.4cm}\\
         & & & & $0.99$ & &  & & $1.41$ & \hspace{0.4cm}\\
         & & & & $1.01$ & &  & & $0.65$ & \hspace{0.4cm}\\
         & & & & $1.20$ & &  & & $0.19$ & \hspace{0.4cm}\\
        \hline
        Figure & \hspace{0.4cm} & \multicolumn{1}{c}{Spacetime} & \hspace{0.4cm} & \multicolumn{1}{c}{$a/M$} & \hspace{0.4cm} & \multicolumn{1}{c}{$k/M$} & \hspace{0.4cm} & \multicolumn{1}{c}{$r_{ISCO}/M$} & \hspace{0.4cm}\\
        \hline
        Fig.~\ref{fig. compare kerr} & & Ghosh & & $0.50M$ & & $0.10$ & & $3.85$ & \hspace{0.4cm}\\
         & & & &  & & $0.20$ & & $3.42$ & \hspace{0.4cm}\\
         & & & & $0.70M$ & & $0.10$ & & $2.92$ & \hspace{0.4cm}\\
         & & & &  & & $0.20$ & & $2.33$ & \hspace{0.4cm}\\ 
         & & Kerr & & $0.50M$ & & - & & $4.23$ & \hspace{0.4cm}\\
         & & & & $0.70M$ & & - & & $3.39$ & \hspace{0.4cm}\\
        \hline
    \end{tabular}
    \caption{The value of ISCO for each configuration we tested.}
    \label{tab. ISCO}
\end{table}

\begin{figure}[htbp!]
    \centering
\includegraphics[width=1\linewidth]{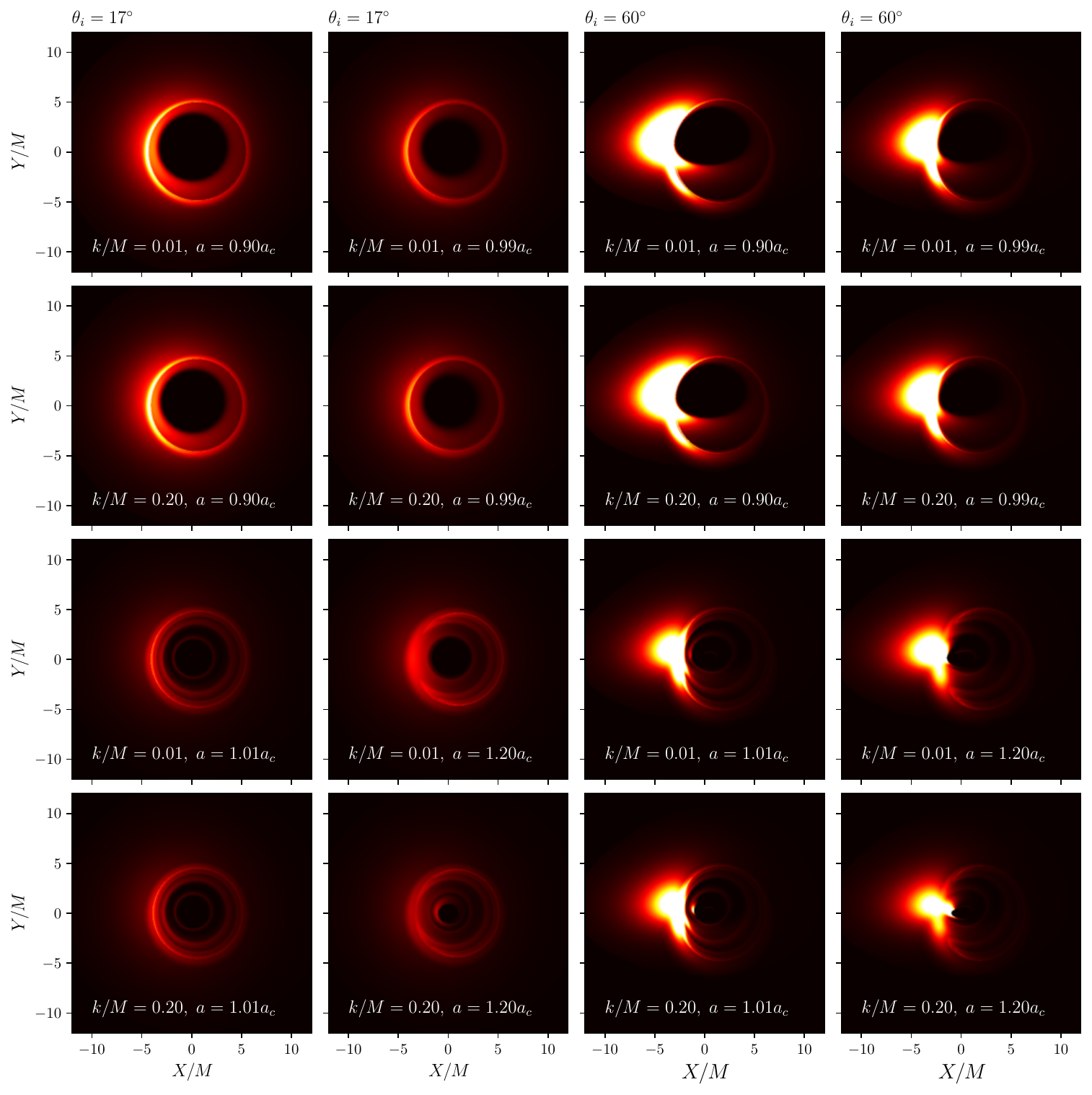}
    \caption{Appearance of the Ghosh RBH (first and second rows) and its superspinar counterparts (third and fourth rows) surrounded by a thin accretion disk for inclination angles $\theta_i = 17^{\circ}$ (first and second columns) and $\theta_i=60^\circ$ (third and fourth columns). The images are normalized to the same maximum intensity.}
    \label{fig. 03rad redshifted images}
\end{figure}

\begin{figure}[htbp!]
    \centering
\includegraphics[width=1\linewidth]{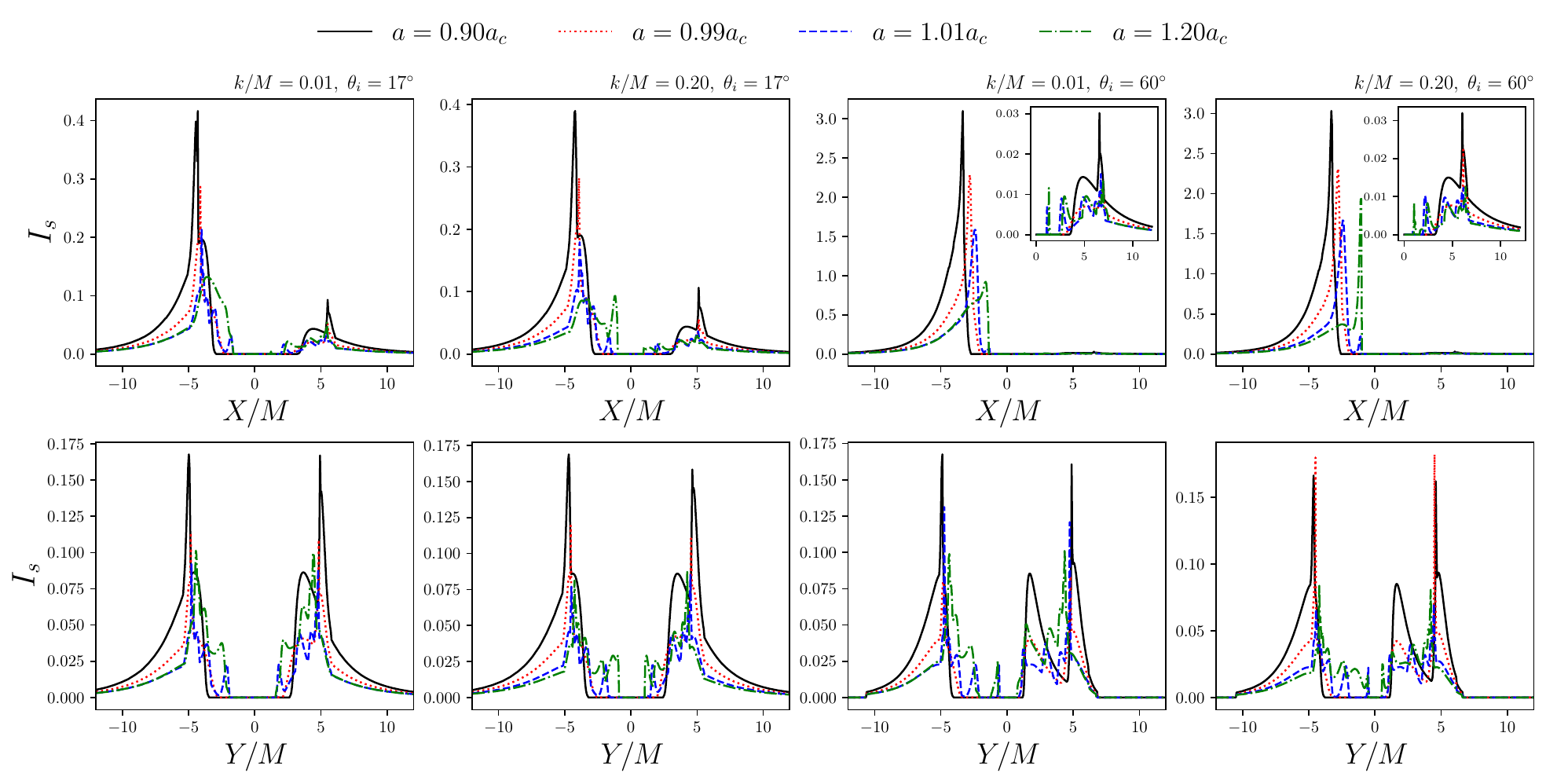}
    \caption{Intensity cross-section corresponding to the central region of the $x$-axis (top row) and $y$-axis (bottom row) for each images presented in Fig.~\ref{fig. 03rad redshifted images}. The sharp peaks indicate the locations of photon ring emissions.}
    \label{fig. intensity cross section ghosh}
\end{figure}

\begin{figure}[htbp!]
    \centering
\includegraphics[width=1\linewidth]{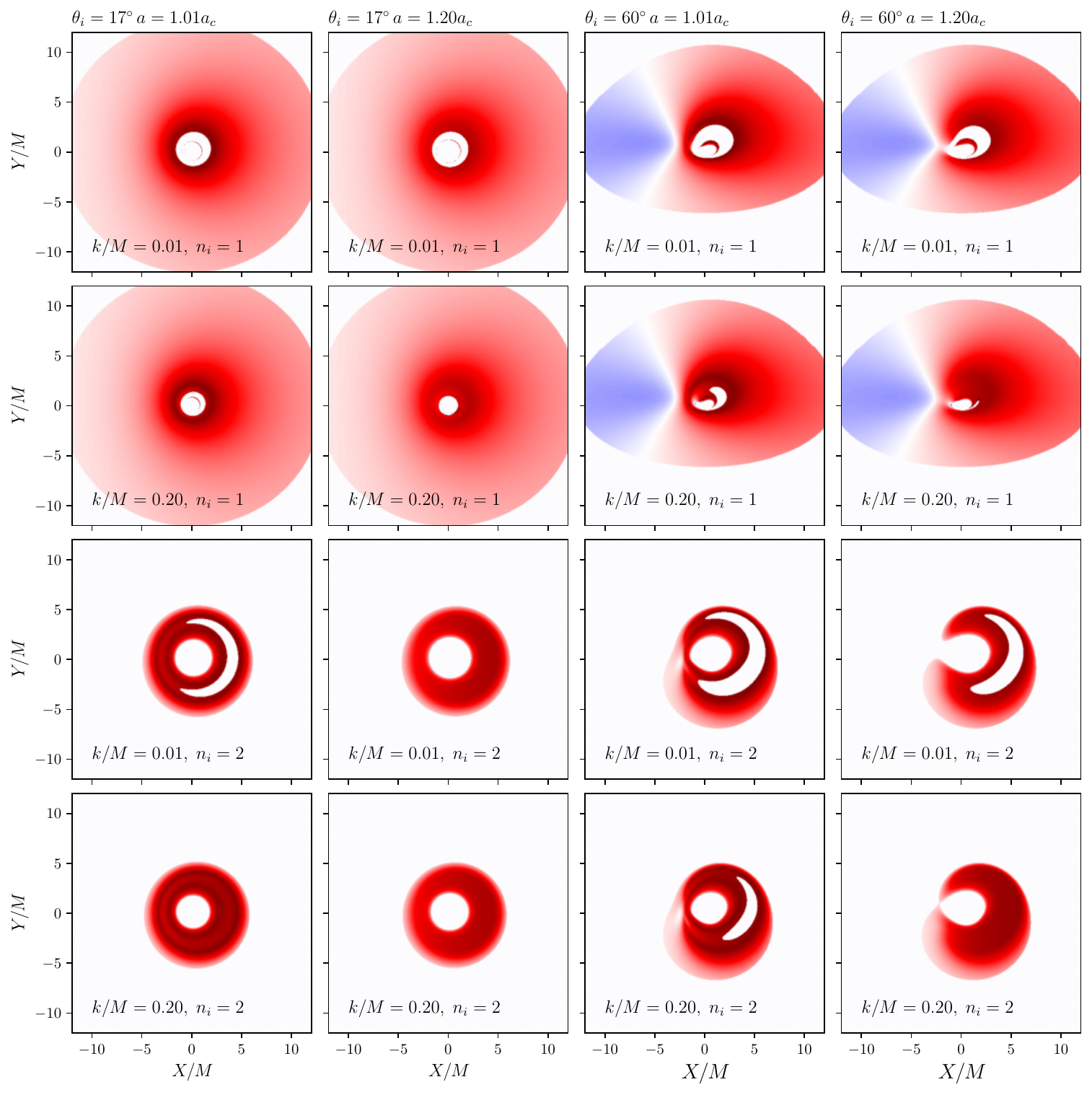}
    \caption{Redshift factor $\tilde{g}$ in the superspinar configurations for the intersection number $n_i=1$ (first and second rows) and $n_i=2$ (third and fourth rows). The white region indicates zero redshift, i.e. $\tilde{g}=1$ at the corresponding number of intersection.}
    \label{fig. redshift profile}
\end{figure}

We generate images of both BH and superspinar configurations by varying the parameter $k$, as shown in Fig.~\ref{fig. 03rad redshifted images}. The intensity cross-sections along the midpoints of the $x$-axis and $y$-axis of the images are also presented in Fig.~\ref{fig. intensity cross section ghosh}. Specifically, we choose $k/M\in\{0.01,0.20\}$ with $a/a_c\in\{0.90,0.99,1.01,1.20\}$, so that one can investigate the effect of the free parameter at its minimum value and at a slightly higher value, as well as the effect of spin near the extremal configuration and beyond. The ISCO value for both configurations are presented in Table~\ref{tab. ISCO}. In the BH case, the image features generally resemble those of a typical rotating BH spacetime. At low inclination angles, the image consists of an almost perfectly circular dark patch, surrounded by a ring-shaped intensity accumulation, which corresponds to the secondary image. At higher angles, the dark patch becomes more distorted, forming a deformed circular shadow encircled by a thin ring. This thin ring is also present at low angles but overlaps with the accumulated intensity of the secondary image. As the inclination increases, the thin ring slightly separates from the dark region, and becomes the boundary of the deformed dark patch.

In contrast, the superspinar configuration exhibits distinctly different image features. At low inclination angles with $a=1.01a_c$, multiple visible ring-like secondary images appear, whereas for $a=1.20a_c$, a crescent-shaped secondary image is formed. At higher inclination angles, these rings become dimmer and less prominent. We also observe that the free parameter $k$ does not significantly affect the appearance in the superspinar configuration with slightly higher spin above the critical limit. However, as the spin increases, the appearance changes dramatically. A larger value of $k$ leads to a bright peak intensity near the center of the image. This is primarily due to the much lower ISCO value for $k/M = 0.20$ compared to $k/M = 0.01$, which, according to our accretion disk model, causes the peak intensity to shift closer to the center.

In addition to the appearance images, we analyze the redshift behavior shown in Fig.~\ref{fig. redshift profile}. We plot the redshift factor $\tilde{g}$ detected from the surrounding accretion disk, which extends from $0.9R_{ISCO}$ to $12M$. Since the redshift behavior of the Ghosh BH exhibits generic features similar to those found in other BH spacetimes (see, for instance, Ref.~\cite{Hou:2022eev}), we focus here only on the redshift profile in the superspinar state.

It can be seen that lower spin values result in more significant redshift effects in the inner region of the disk. An interference-like pattern appears for $a=1.01a_c$ and $\theta_i=17^\circ$, resembling the ring structures observed in Fig.~\ref{fig. 03rad redshifted images}. The blueshift due to Doppler effects only appears at high inclination angles ($\theta_i=60^\circ$) on the first intersection ($n_i=1$). We observe no significant distinction arising from variations in the parameter $k$, apart from differences in the ISCO radius and slight different in the redshift values. The distinct features near the center for $k/M=0.01$ and $k/M=0.2$ are primarily due to the difference in ISCO radii; a higher $k$ value leads to a smaller ISCO radius, allowing the disk to extend deeper into the core region. We therefore conclude that the discrepancies between the images for different values of $k$, especially at high spins, are mainly due to the differences in the inner radius\textemdash and consequently the intensity profile\textemdash of the accretion disk.

\begin{figure}[htbp!]
    \centering
\includegraphics[width=1\linewidth]{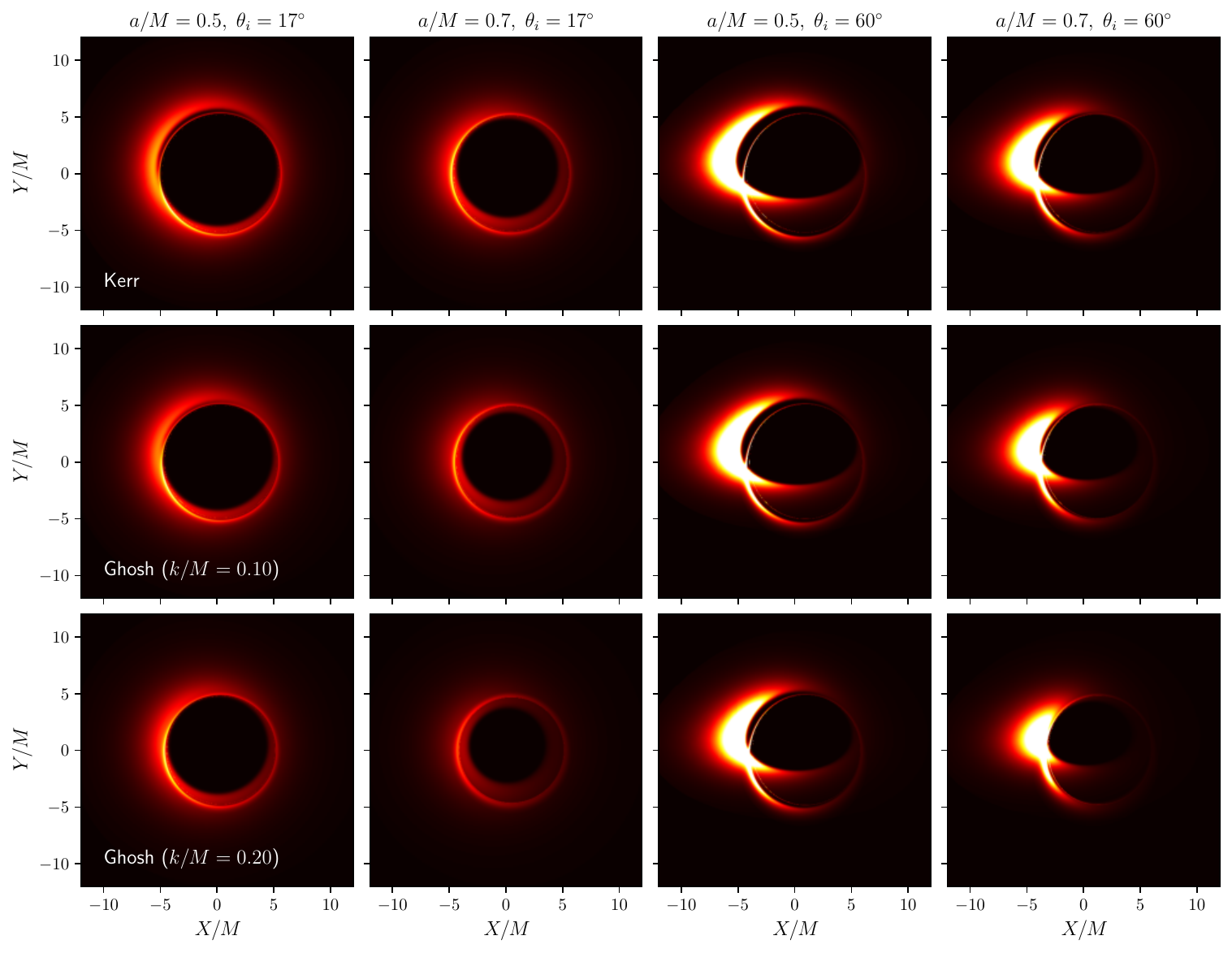}
    \caption{Image comparison between the Kerr (first row) and the Ghosh (second and third rows) BHs with the same spin parameter in each column, for inclination angles $\theta_i = 17^{\circ}$ (first and second column) and $\theta_i=60^\circ$ (third and fourth column). The intensity of each image is normalized to the same maximum intensity.}
    \label{fig. compare kerr}
\end{figure}

These superspinar images differ significantly from those presented in other studies, such as Refs.~\cite{Lamy:2018zvj} and~\cite{Eichhorn:2022bbn}. This discrepancy mainly stems from the different intensity profiles used for the accretion disk and the location at which the emission ceases. In our approach, the disk emission profile stops slightly below ISCO, determined via Eq.\eqref{eq. determine ISCO}. On the other hand, Ref.~\cite{Eichhorn:2022bbn} impose a relatively sharp cutoff at $4M$ and $5M$, repsectively. This choice significantly affects the appearance of secondary images, making the ring-like features more prominent in their results compared to ours. Our approach accounts for redshift effects, which appear to be absent in Refs.~\cite{Eichhorn:2022bbn}. Notably, the inclusion of Doppler redshift introduces asymmetry in the observed intensity, making one side of the image appear brighter than the other. This asymmetry serves as a distinguishing feature of models that incorporate relativistic redshift effects.

\begin{figure}[htbp!]
    \centering
\includegraphics[width=1\linewidth]{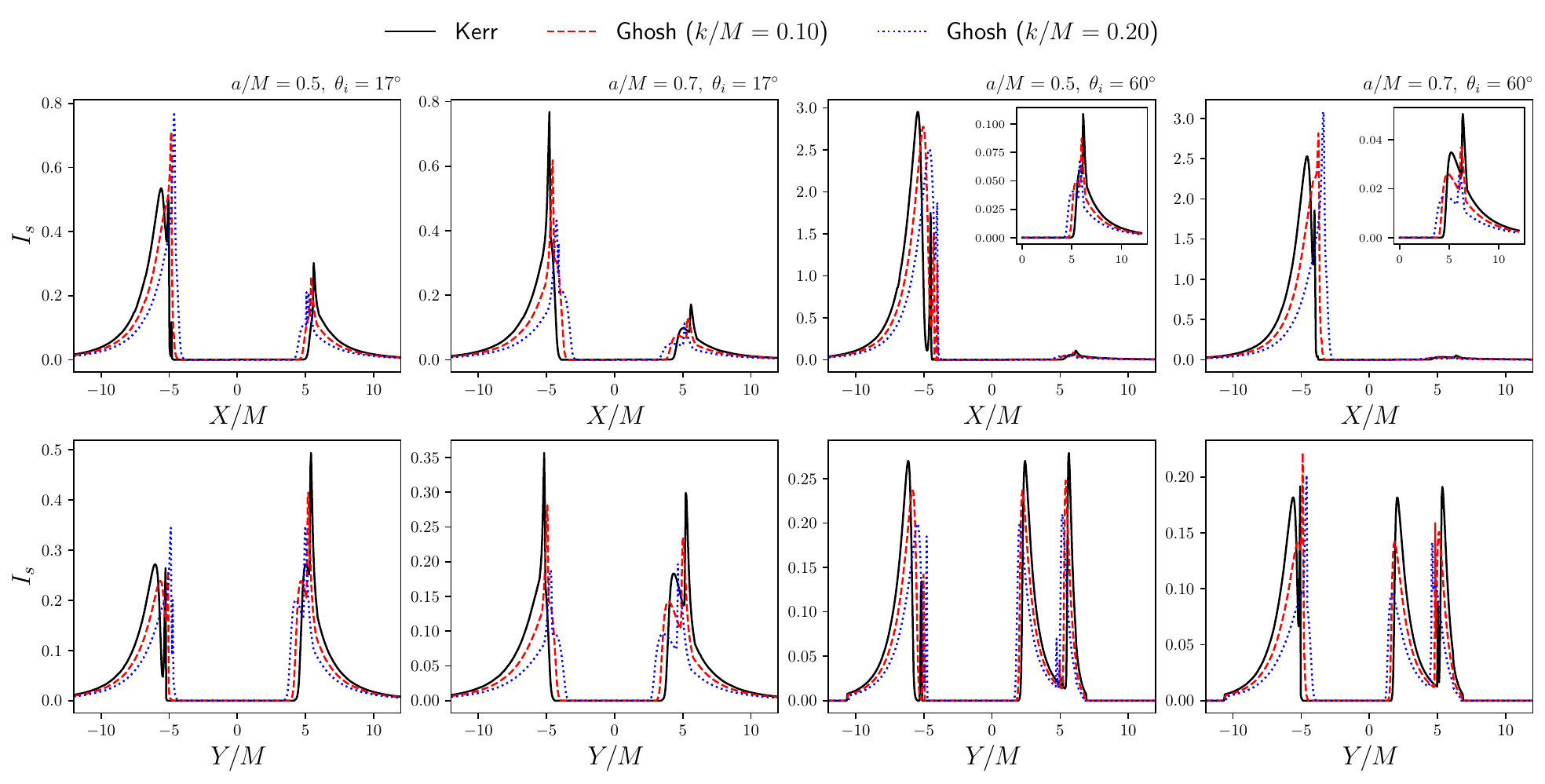}
    \caption{Intensity cross-section corresponding to the central region of the $x$-axis (top row) and $y$-axis (bottom row) for each images presented in Fig.~\ref{fig. compare kerr}. The sharp peaks indicate the locations of photon ring emissions.}
    \label{fig. intensity cross section}
\end{figure}

Finally, to compare the image features of the Ghosh BH with its singular counterpart (the Kerr BH) we analyze how the parameter $k$ influences the appearance of a BH surrounded by a thin accretion disk. The resulting images are shown in Fig.~\ref{fig. compare kerr}, while the intensity cross-sections along both the mid $x$-axis and $y$-axis are presented in Fig.~\ref{fig. intensity cross section} for a quantitative comparison of their differences.

It is clear that the additional parameter $k$ affects both the sizes of the ISCO and the photon ring radius of the BH. In our case, both decrease as $k$ increases, with the ISCO being more strongly affected. This causes the secondary photon ring image to overlap with the direct emission from the accretion disk at larger $k$ values, as shown in the first column of Fig.~\ref{fig. compare kerr}. For $a/M = 0.5$ and $\theta_i = 17^\circ$, the photon ring in Kerr spacetime is clearly separated in the upper portion of the image. In contrast, for Ghosh spacetime with \(k/M = 0.20\), the photon ring overlaps with the inner direct emission from the accretion disk, specifically at the ISCO. Upon examining the intensity cross-section in Fig.~\ref{fig. intensity cross section}, it is evident that the intensity also decreases for smaller ISCO radii. This reduction is primarily attributed to the gravitational redshift effect, which, analogous to the spherically symmetric spacetime, becomes more significant near the horizon, as the redshift factor is given by $\tilde{g} = \sqrt{-g_{tt}}$ with static fluids.

\section{Imaging the black hole's destruction}
\label{sec. image transition}

Let us review the destruction scenario proposed in our previous discussion. Consider an extremal rotating RBH ($a = a_c$) surrounded by a thin accretion disk. Suppose a collapsing null shell at $u=u_c$ with a particular angular momentum causes the BH to gain an additional angular momentum $\delta$, so that $a=a_c \to a_c+\delta$, becoming a regular superspinar. This transition destroys the event horizon, allowing light to pass through the core of the BH.

In Ref.~\cite{Eichhorn:2022bbn}, it is shown that overspinning a RBH leads to an instantaneous \textit{lighting-up} effect, when the secondary image on the central region appears instantaneously as the spin of the BH exceed its critical limit. This occurs because the event horizon is resolved globally, allowing light to pass through the central region as soon as the spin parameter exceeds the critical value $a_c$, even for an arbitrarily small increase $\delta$.

\begin{figure}[htbp!]
    \centering
\includegraphics[width=1\linewidth]{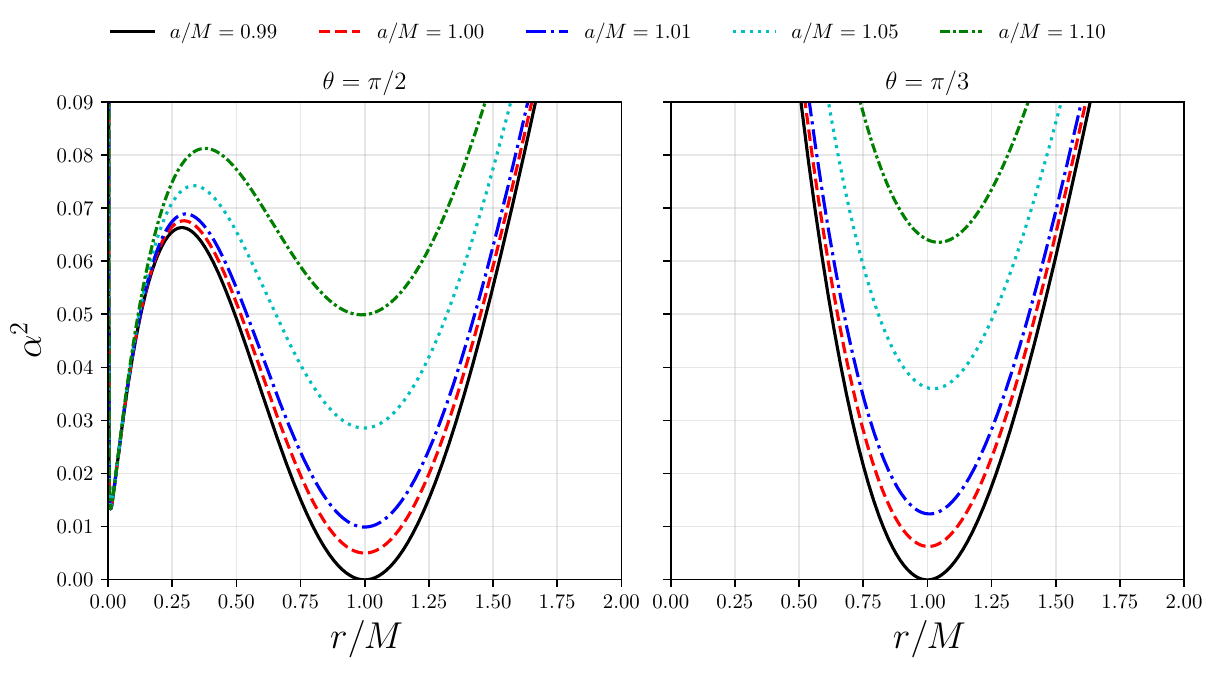}
    \caption{The lapse function $\alpha^2$ on $\theta=\pi/2$ (left) and $\theta=\pi/3$ (right) with $k/M=0.01$ for several values of $a$. The black solid line correspond to the extremal BH configuration.}
    \label{fig. alpha graph}
\end{figure}

The time measured by the observer as a light travels from $\tau_0$ to $\tau_f$ is given by~\cite{Bacchini:2018zom,Chen:2024ibc}
\begin{equation}
    \Delta t = \int^{\tau_f}_{\tau_0} k^t d\tau= \int^{\tau_f}_{\tau_0} \sqrt{\frac{\gamma^{ij}k_i k_j}{\alpha^2}} d\tau,
    \label{eq. delta t}
\end{equation}
where $k^i$ represents the null 4-velocity, and $\tau$ is an affine parameter. It can be seen that the time delay is most significant at the minimum value of $\alpha$ on the equatorial plane (see Fig.~\ref{fig. alpha graph}). If light emitted from the accretion disk undergoes multiple orbits near this minimum region before escaping to infinity, the travel time observed by a distant observer will be significantly longer. This may results in the delay of the formation of the secondary images. Based on this simple reasoning, we expect it is possible to observe a delayed lighting-up of a the transition from a rotating RBH to a superspinar, which is the main idea of this paper. It is important to note that in this scenario, the gradual change in the observed image is not due to a continuous variation in the spin parameter $a$ as discussed in~\cite{Eichhorn:2022bbn}, but rather due to the progression of time $t$.

In our following discussion, we will use these configuration: we set $k/M=0.01$ to maximize the spin parameter, which corresponds to $a_c/M\approx 0.9899$. The spin transition rate is close to instantaneous, in which we choose $\sigma_s=10^{-4}$. 

\subsection{Photon worldlines}

\begin{figure}[htbp!]
    \centering
\includegraphics[width=0.45\linewidth]{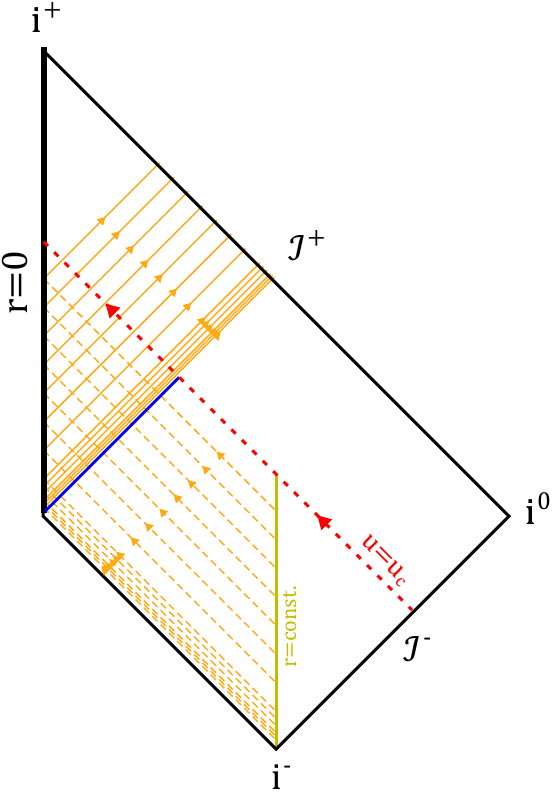}
    \caption{Penrose diagram sketch of radially ingoing null trajectories (dashed orange lines) originating from $r=\text{const.}$ at $u<u_c$, which propagate toward the BH and reflect at $r=0$, continuing as radially outgoing null trajectories (solid orange lines).}
    \label{fig. penrose diagram null trajectories}
\end{figure}

Based on the spacetime diagram in Fig.~\ref{fig. penrose diagram destroying BH}, we can qualitatively infer the behavior of radially moving light rays in the dynamic spacetime. In Fig.~\ref{fig. penrose diagram null trajectories}, we sketch the worldlines of radially ingoing light rays in the extremal BH scenario. Instead of reaching the singularity, these rays ‘bounce’ at the center $r=0$ and continue as outgoing light rays\footnote{Strictly speaking, the center $r=0$ may not correspond to a straight line but rather a curved one when calculated properly, as discussed in Ref.~\cite{Frolov:2016gwl}. Nevertheless, the present sketch serves to illustrate the essential idea.}. After the null shell destroys the BH at $u=u_c$, all of these photons can escape to null infinity. Furthermore, photons arriving from past infinity accumulate near the extremal horizon, which results in a sudden burst of radiation at the destruction event: an effectively infinite number of photons, previously piled up close to the extremal horizon, escape within nearly the same outgoing null coordinate.

\begin{figure}[htbp!]
    \centering
\includegraphics[width=1\linewidth]{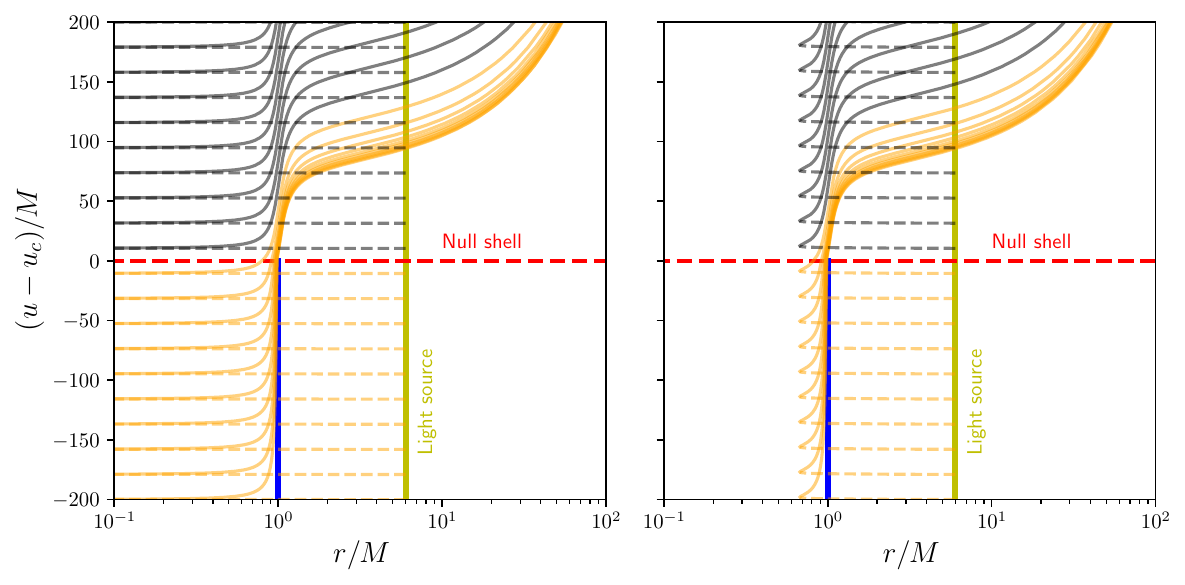}
    \caption{Radial (left) and non-radial (right) ingoing null trajectories (dashed orange and gray lines) originating from $r=\text{const.}$ propagate toward the BH and reflect at $r=0$, continuing as outgoing null trajectories (solid orange and gray lines). The orange and gray curves correspond to light rays sourced from $u<u_c$ and $u>u_c$, respectively.}
    \label{fig. spacetime diagram null}
\end{figure}

We extend our analysis using a slightly different form of spacetime diagram, similar to those presented in Refs.~\cite{Frolov:2016gwl,Frolov:2017rjz}. In Fig.~\ref{fig. spacetime diagram null}, we place a light source at the equatorial plane at $r=6M$, which emits both ingoing radial ($d\varphi/d\tau=0$, $d\theta/d\tau=0$) and non-radial ($d\varphi/d\tau\neq0$, $d\theta/d\tau=0$) light rays confined to the equatorial plane over a finite interval of $u$. The diagram shows that all photons crossing the BH eventually accumulate at the (inner) horizon. These results confirm the qualitative picture suggested by our sketch analysis using the Penrose diagram in Fig.~\ref{fig. penrose diagram null trajectories}.

This simple analysis suggests that one would observe a flash of light at the moment of the BH's destruction, with no gradual transition in the observational image as light rays instantly expose the BH's core. However, our ray tracing results reveal a different outcome, and we investigate the underlying reason in the following section.

\subsection{Required method for gradual image transition}
\label{sec. essential technique pinhole}

There is a key requirement for obtaining the image transition of BH destruction, despite the fact that all outgoing photons from the dissolved horizon are released at a single instant. This requirement is tied to the imaging procedure: in the pinhole ray tracing technique, incident light rays must travel \textit{exactly} through the pinhole. Consequently, radially directed photons (photons with \textit{zero} impact parameter with respect to $r=0$) are captured at the center of the image frame, as illustrated by the green line in Fig.~\ref{fig. observer illustration}. This condition plays a crucial role in producing the observed image transition.

\begin{figure}[htbp!]
    \centering
\includegraphics[width=0.7\linewidth]{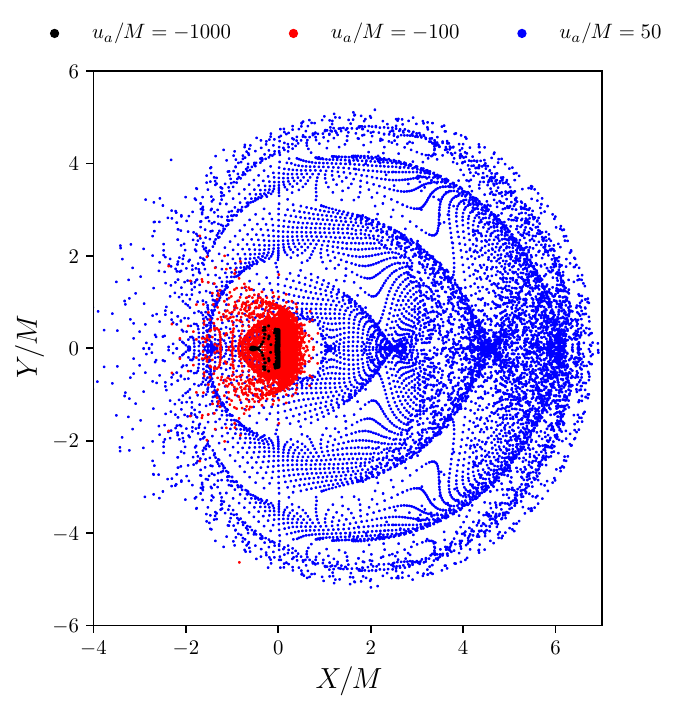}
    \caption{Detected photon locations on the image frame positioned at $r_\infty\sim500M$ for photons that escaped the BH horizon. The photons emitted at $u_a=-1000M$ (black points), $u_a=-100M$ (red points), and $u_a=50M$ (blue points), from a source located at $r_a=10M$ on the equatorial plane, pointed towards the BH horizon.}
    \label{fig. light accretion disk}
\end{figure}

Consider a light source located in the accretion disk plane (at $r=r_a$ and $\theta=\pi/2$) that emits a bundle of photons in different directions toward the BH horizon at $u=u_a$. We then record their relative velocities at spatial infinity, chosen as $r_\infty \sim 500M$, and map them onto the image frame coordinates $(X,Y)$ corresponding to the detected photons. With this method, one can identify the regions of the image frame that capture outgoing light rays from the destroyed horizon, which were originally emitted from the accretion disk plane.

In Fig.~\ref{fig. light accretion disk}, we present a set of sample points indicating which regions of the image frame capture the light rays emitted from $r_a = 10M$. It can be seen that ingoing light rays crossing the horizon, emitted at $u_a \sim -1000M$, reach only the central part of the image once they emerge from the destroyed horizon. This demonstrates that photons released at the moment of BH destruction propagate radially at spatial infinity. As $u_a$ increases, the covered area of the image grows, reaching its maximum once $u_a > u_c$, when the horizon no longer exists. Since light rays emitted at earlier $u$ emerge from the destroyed horizon ahead of those from later $u$, this implies that, for a brief moment, all such rays are localized at the center of the image frame before gradually spreading outward over time. This mechanism underlies the gradual image transition obtained with the pinhole ray tracing technique.

However, this situation may not be directly applicable to actual observations. The pinhole ray tracing technique implicitly assumes an instrument with infinite resolution. In reality, one would expect that finite-resolution imaging would observe a sudden flash of light from the previously trapped rays, possibly followed, at best, by a gradual transition. These aspects may be of future interest in BH destruction imaging to obtain more accurate predictions of potential observables.

\subsection{Image, horizon's shadow, and overall intensity transition}

The BH configuration and destruction scenario are chosen as follows. To see how the spin change affect the transition, we choose $\delta/M \in \{5\times10^{-3},1\times10^{-2},5\times10^{-2}\}$. Our previous investigation revealed that the most distinct differences in appearance and intensity between the BH and superspinar configurations occur at low inclination angles, as shown in Fig.~\ref{fig. 03rad redshifted images}. Therefore, we only consider image transitions at an inclination of approximately $17^\circ$.

\begin{figure}[htbp!]
    \centering
\includegraphics[width=1\linewidth]{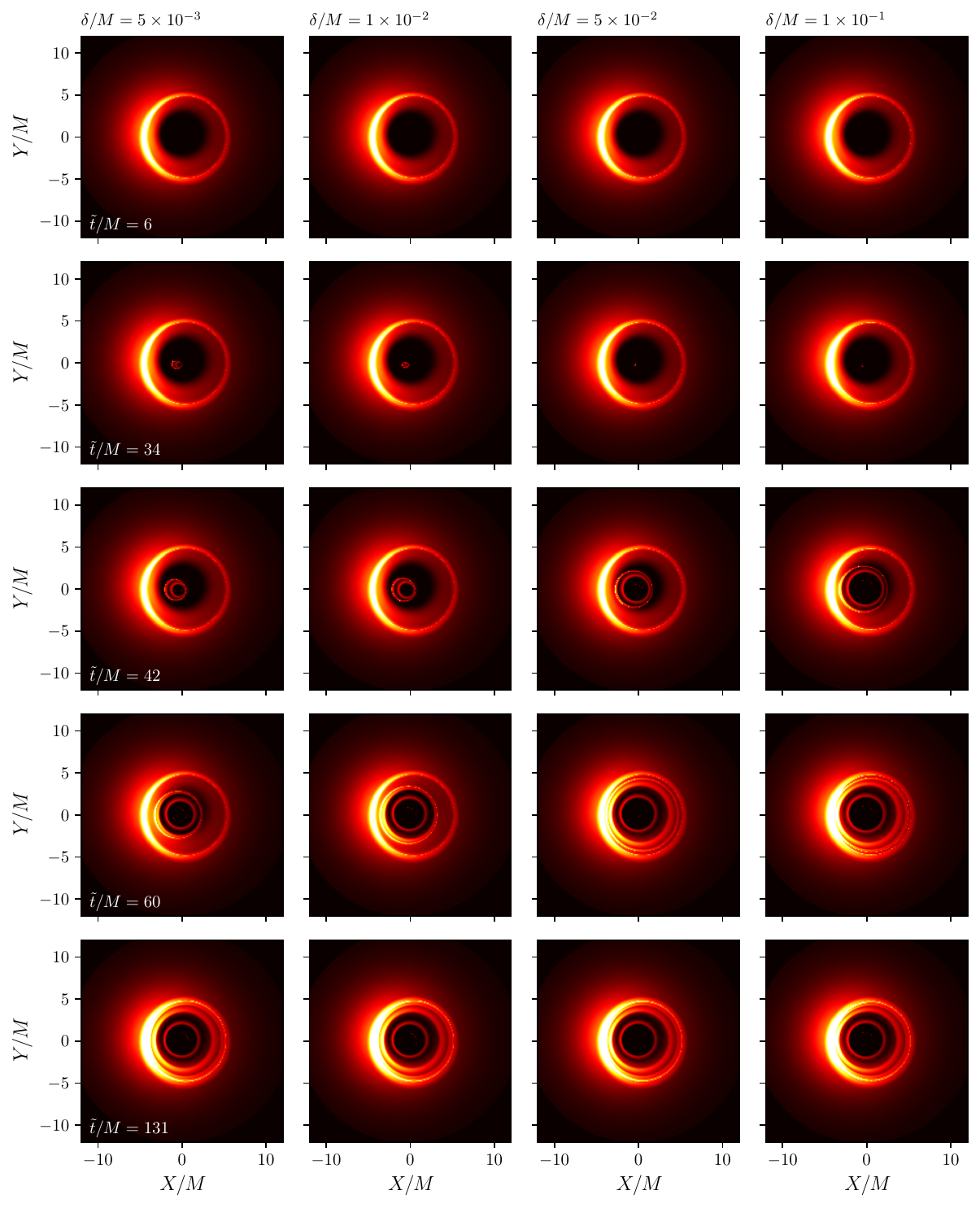}
    \caption{The appearance evolution of the recently destroyed extremal Ghosh BH ($a = a_c$) over time into its superspinar counterparts (\(a = a_c + \delta\)) with an inclination angle of $\theta_i = 17^{\circ}$ for (left to right) $\delta/M \in\{ 5\times10^{-3},1\times10^{-2},5\times10^{-2},1\times10^{-1}\}$.} 
    \label{fig. destroying bh}
\end{figure}

\begin{figure}[htbp!]
    \centering
\includegraphics[width=1\linewidth]{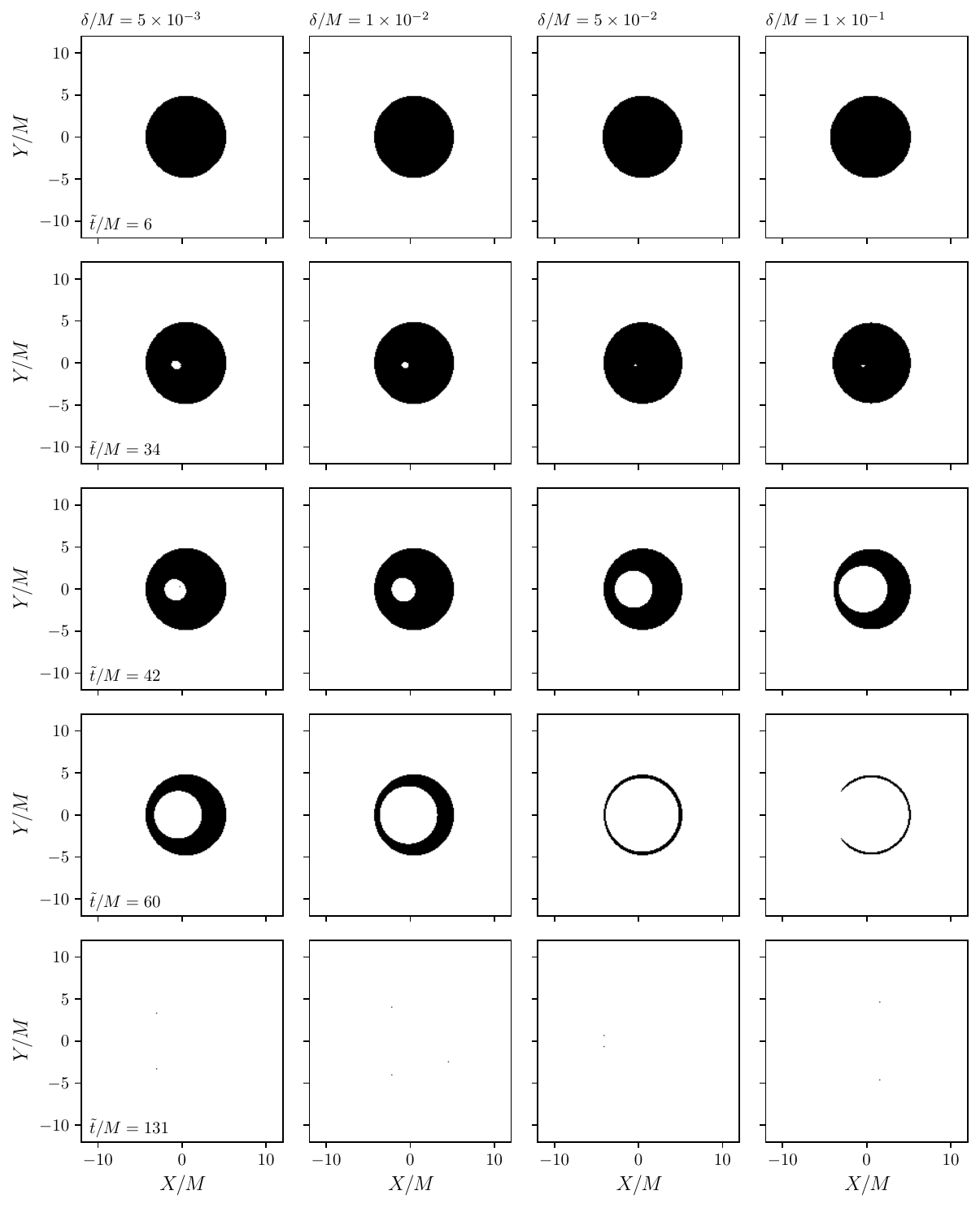}
    \caption{The horizon's shadow evolution of the recently destroyed extremal Ghosh BH (\(a = a_c\)) over time into its superspinar counterparts (\(a = a_c + \delta\)) with an inclination angle of \(\theta_i = 17^{\circ}\) for (left to right) $\delta/M \in\{ 5\times10^{-3},1\times10^{-2},5\times10^{-2},1\times10^{-1}\}$.} 
    \label{fig. horizon shadow}
\end{figure}

To image the transition, we perform the ray tracing procedure over a range of $t_{(0)}$, from the initial observation time $t_{(0)}^i$ to the final time $t_{(0)}^f$, with a uniform interval. The total observation length is chosen to be $t_{(0)}^f - t_{(0)}^i = 200M$ with an interval of $2M$, so that 100 images are generated for each scenario. The value of $t_{(0)}^i$ depends on $\delta$, which is determined through trial and error to obtain the observation closest to the destruction moment for each $\delta$. Since the observed destruction time varies, we introduce a new variable called the \textit{shifted time} $\tilde{t}$, where we shift the time for each scenario so that all of their destruction events\textemdash denoted by the beginning of first photon ring formation\textemdash are observed at approximately the same $\tilde{t}$.

We also impose a simplifying assumption regarding the matter responsible for the emission. The intensity profile, ISCO, and accretion flow of the disk are assumed to remain unchanged throughout the entire process: we consistently adopt the same disk characteristics as those in the extremal BH state, even though the collapsing null shell may significantly affect the surrounding dynamics. Furthermore, the destruction of the horizon would expose matter that was previously trapped inside. However, since we expect a strong redshift in the inner region of the superspinar (see Fig.~\ref{fig. redshift profile}), which would substantially dim the observed intensity, we neglect the contribution of light emitted from such sources. A proper treatment accounting for dynamical emitting matter would require a more comprehensive analysis, which lies beyond the scope of this work.

\begin{figure}[htbp!]
    \centering
\includegraphics[width=1\linewidth]{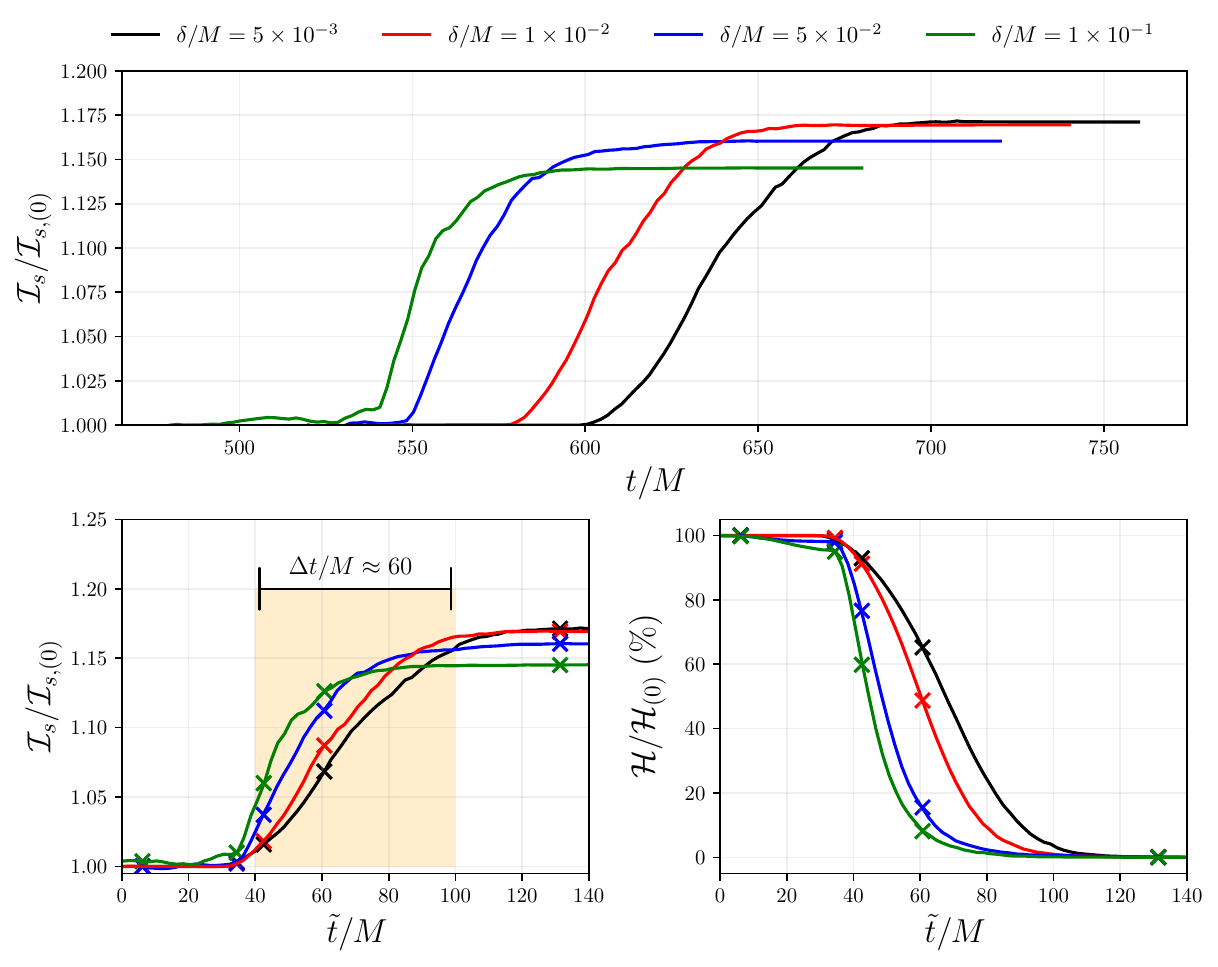}
    \caption{(Top) Evolution of the integrated intensity $\mathcal{I}_s$ over time $t$ normalized to its initial intensity at $\tilde{t}/M=0$ ($\mathcal{I}_{s,(0)}$). (Bottom left) Same as the top figure with shifted time $\tilde{t}$. (Bottom right) Evolution of the percentage of horizon's shadow area $\mathcal{H}$ over shifted time $\tilde{t}$ relative to its horizon's shadow area at $\tilde{t}/M=0$ ($\mathcal{H}_{(0)}$). The cross points on the bottom figure indicates the moment of each value shown in Figs.~\ref{fig. destroying bh} and~\ref{fig. horizon shadow}.}
    \label{fig. intensity destroying bh}
\end{figure}

We generate the image transition along with the horizon's shadow as observed by a distant observer. To obtain a quantitative result, we also plot the temporal evolution of both the integrated intensity $\mathcal{I}_s$ and the relative horizon shadow area $\mathcal{H}$,
\begin{equation}
    \mathcal{I}_s=\sum_{x_p,y_p}I_s(x_p,y_p),\qquad \mathcal{H}=\sum_{x_p,y_p} H_A(x_p,y_p),
\end{equation}
where $H_A(x_p,y_p)=1$ if the computed light rays corresponding to the $(x_p,y_p)$ pixel remain outside but close to the horizon with outgoing motion at $u \ll u_c$ (as will be shown shortly), and $H_A(x_p,y_p)=0$ otherwise. The resulting transition images, horizon's shadow, and temporal evolution of both $\mathcal{I}_s$ and $\mathcal{H}$ are shown in Figs.~\ref{fig. destroying bh},~\ref{fig. horizon shadow}, and~\ref{fig. intensity destroying bh}, respectively.

Judging from the generated images in Fig.~\ref{fig. destroying bh}, we observe that all superspinning scenarios exhibit a similar appearance transition: the secondary images expand outward from the center until they form final states resembling the superspinar images shown in Fig.~\ref{fig. 03rad redshifted images}. This behavior arises primarily because the horizon shadows in Fig.~\ref{fig. horizon shadow} are \textit{destroyed} in a similar manner, dissolving outward from the central region. In addition, the dissolving horizon tends to shift leftward, disappearing more rapidly on the left side than on the right, which directly affects the resulting transition images. Although a more detailed analysis of the null geodesics is required to fully understand this behavior, it is likely caused by frame-dragging effects from the BH spin.

\begin{figure}[htbp!]
    \centering
\includegraphics[width=1\linewidth]{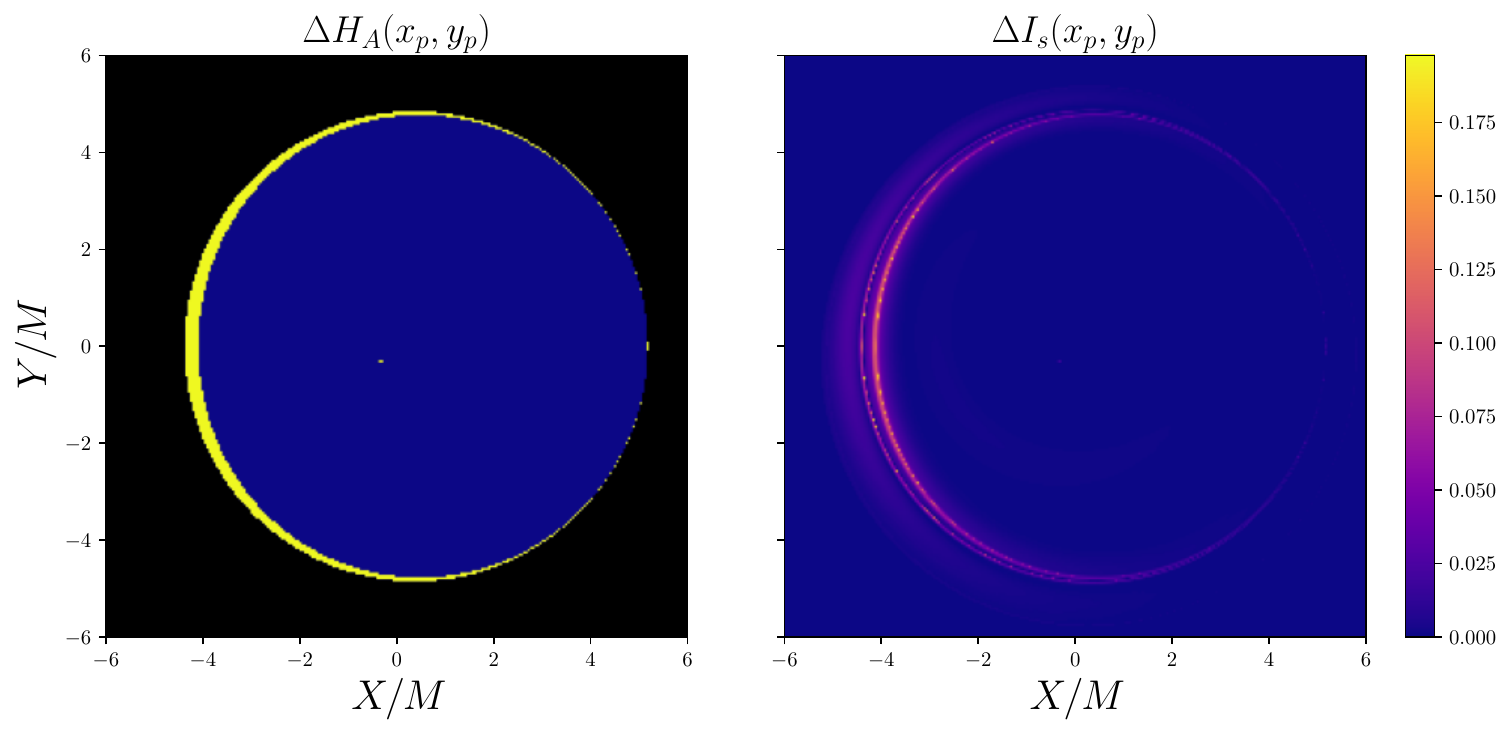}
    \caption{(Left) Horizon's shadow shape and (Right) appearance intensity differences between observations at $t/M=400$ and $t/M=480$ with $\delta/M=10^{-1}$. In the left figure, the yellow area depicts the observed horizon's shadow at $t/M=400$, while the blue area represent the intersection between the horizon's shadow at $t/M=480$ and $t/M=480$.}
    \label{fig. horizon shape difference}
\end{figure}

We perform a quantitative comparison of each scenario by examining the temporal evolution of the integrated intensity and horizon area shown in Fig.~\ref{fig. intensity destroying bh}. Our results indicate that, although the destruction events occur at the same instant $u_c$, the observable duration of the transition—from the initial appearance of the secondary ring to the final state—takes place later in time for smaller $\delta$. When measured in shifted time, the transition is also longer for smaller $\delta$, primarily due to the smaller minimum value of the lapse function in the superspinar final state (see Fig.~\ref{fig. alpha graph}). Furthermore, we find that the observed transition duration $\Delta t$ is roughly within $\Delta t/M \approx 60$, and could be longer for smaller values of $\delta/M$.

An interesting transition occurs for $\delta/M = 10^{-1}$. Owing to the large spin difference, noticeable changes arise in both the shape of the horizon shadow and the image appearance prior to destruction (see Fig.~\ref{fig. horizon shape difference}). The shadow area becomes ``compressed'' from the left side until the horizon destruction sets in. Consequently, the images appear slightly brighter just before the first ring emerges (see the green curves in Fig.~\ref{fig. intensity destroying bh}).

\begin{figure}[htbp!]
    \centering
\includegraphics[width=0.7\linewidth]{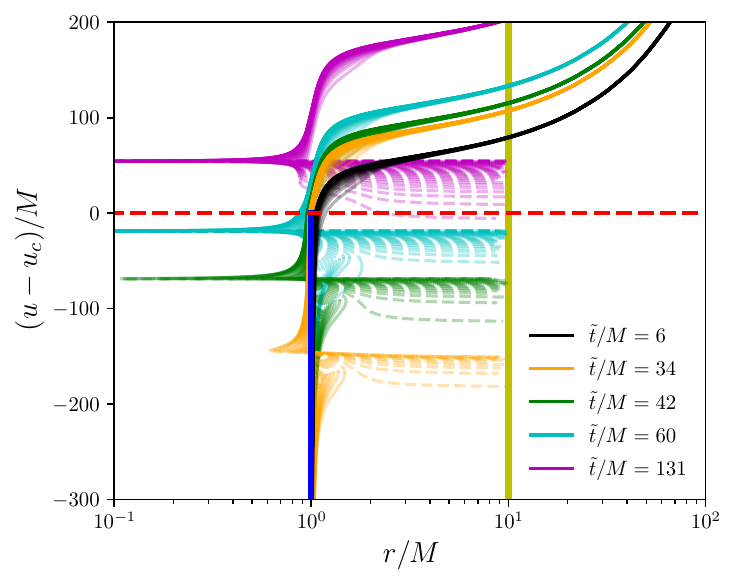}
    \caption{Null trajectories observed at the chosen $\tilde{t}$, which carry the accretion disk's intensity $I_s(x_p,y_p)$ and horizon area $H_A(x_p,y_p)$ information to the region $Y/M=0$ and $-4<X/M<0$ in Fig.~\ref{fig. destroying bh}, for the case $\delta/M=5\times10^{-3}$.} 
    \label{fig. trajectories tilde}
\end{figure}

To illustrate how the secondary images are revealed, we perform a calculation analogous to that in Fig.~\ref{fig. spacetime diagram null}, tracing null geodesics backward from the observer toward the BH at a chosen observer time. The result is shown in Fig.~\ref{fig. trajectories tilde} for the case $\delta/M=5\times10^{-3}$. These null geodesics correspond to photons detected at $Y/M=0$ and $-4\leq X/M\leq0$ in Fig.~\ref{fig. destroying bh}. Since the light rays are integrated backward, we may instead interpret them as if they were sent from the observer toward the BH in ‘forward motion’—thus, in this discussion, the photons should be viewed as \textit{coming} from top to bottom.

Different trajectory behaviors are observed at different $\tilde{t}$. For $\tilde{t}=6$, all light rays never actually reach the horizon: they cross $u_c$ above the horizon location $r_H$ and remain in outgoing motion. As noted, these rays result in $H_A(x_p,y_p)=1$. In contrast, at $\tilde{t}=34$, $\tilde{t}=42$, and $\tilde{t}=60$, the light rays can be grouped into three categories: (i) those that reach $r>r_H$ at $u=u_c$, (ii) those that reach $r<r_H$ at $u=u_c$ but are pulled back toward the horizon, and (iii) those that reach $r<r_H$ at $u=u_c$ and then escape to infinity. Category (i) rays behave similarly to those at $\tilde{t}=6$, yielding $H_A(x_p,y_p)=1$ and no recorded intensity. Category (ii) rays also behave in this way; although they briefly penetrate below $r_H$, they eventually reemerge as outgoing rays at $u\ll u_c$, so $H_A(x_p,y_p)=1$, though they may still carry intensity information as they exit the horizon. Finally, category (iii) rays escape to infinity and are thus primarily responsible for carrying the intensity information that produces the image transition, corresponding to $H_A(x_p,y_p)=0$. The fraction of category (iii) rays increases with $\tilde{t}$, which causes the secondary image to widen and the dissolved horizon area to expand. By $\tilde{t}=131$, no rays cross $u_c$ before becoming ingoing rays, allowing all photons to escape to infinity and thus transmitting the entire accretion-disk intensity, giving $H_A(x_p,y_p)=0$.

\subsection{Is the timescale relevant?}

Regardless of the flash of light produced at the moment of BH's destruction, one may ask about the timescale of the image transition: \textit{Is it relevant for observations on Earth?} Capturing a BH image with the EHT requires a substantial observational period. For example, the first image of M87* was reconstructed from data collected over the course of an entire week~\cite{EventHorizonTelescope:2019dse}. If the transition timescale is significantly shorter than the observational window, the transition would be effectively unobservable and thus irrelevant for practical purposes.

Let us consider the transition time $\Delta t/M\approx 60$. Converting the natural units back to SI, $M\to GM/c^2$ and $t\to ct$, we obtain an estimated transition time of $\Delta t=60GM/c^3$. For the M87* BH, with a mass of approximately $6.5\times 10^9 M_{\odot}$~\cite{EventHorizonTelescope:2019dse}, the corresponding transition time is
\begin{equation}
    \Delta t_{M87*} \approx 19.2\times 10^{5}~\text{s} \approx 22.2~\text{days}, 
\end{equation}
roughly three times larger than the observation time of the EHT. However, despite the comparable timescale, this does not necessarily imply that the transition is directly measurable. These results suggest that such a transition might be more relevant for even more massive BHs.

We can estimate the BH mass at which these phenomena become relevant based on current observational data. For instance, if we aim to capture five observational images during the transition and assume a transition time of approximately $\Delta t = 60 \, GM/c^3$, the observation must be conducted within intervals of roughly $\Delta t = 12 \, GM/c^3$. Given the observational time frame of EHT ($\sim 7$ days)~\cite{EventHorizonTelescope:2019dse}, the corresponding relevant BH mass \( M_{obs} \) can be determined, resulting in
\begin{equation}
    M_{obs} \approx 10.2\times10^{9}M_\odot,
\end{equation}
which is less than twice the mass of M87* and remains well below the theoretical maximum for a BH mass~\cite{Kingblackhole}. However, due to the sudden bursts of light, overlapping frames in daily observations, and finite resolutions, the resulting image transition may not be as distinct as depicted in Fig.~\ref{fig. destroying bh}.

\section{Summary}
\label{sec. conclusion}

In this paper, we explored the characteristics of a RBH and its superspinar counterpart, specifically focusing on the Ghosh BH, which is surrounded by a thin accretion disk. We generated images using a ray-tracing method, employing the Hamiltonian formalism to compute the null geodesics within our ray-tracing code. We also calculated the ISCO radius to define the inner boundary of the accretion disk and analyzed the effects of gravitational and Doppler redshift on the accretion flow. Our findings indicate that the Ghosh BH closely resembles the Kerr BH. As the parameter $k$ increases, both the Ghosh BH's ISCO and photon ring radius decrease. In contrast with the BH spacetime, the lack of horizons in the superspinar case results in a distinctly different image, marked by the presence of inner secondary images. However, these secondary images become less visible at higher inclination angles due to strong Doppler effects, which significantly diminish their intensity.

We also examined the image transition of a recently destroyed rotating RBH caused by a collapsing null shell with angular momentum, effectively removing the BH’s horizons and transforming it into a superspinar. The potential production of a sudden flash of light at the moment of BH destruction was discussed, and a method to account for it is incorporated in the pinhole ray tracing technique we employed. The resulting image transitions show that the transition time is sensitive to the additional angular momentum acquired, with an approximate duration of $\Delta t/M \approx 60$ across all scenarios. However, the sudden flash of light may render the gradual transition observed in this study irrelevant for actual observations.

Furthermore, we discussed the significance of the transition time in the context of current observational capabilities, while disregarding the sudden burst of light. For a BH with a mass comparable to M87*, the image transition time is approximately 22 days. The EHT typically requires about one week to capture an image~\cite{EventHorizonTelescope:2019dse}, and therefore, to achieve meaningful observations of the transition, at least five consecutive weeks of monitoring would be necessary. Based on this, we estimate that the image transition becomes more relevant for BHs more massive than M87*, at least less than twice its mass (around $10.2\times10^{9}M_\odot$), which is still well below the theoretical maximum BH mass~\cite{Kingblackhole}.

Finally, we highlight several caveats that could inform future investigations. Studies of BH destruction aimed at testing the validity of the WCCC typically involve test particles or test fields. It may be more relevant to consider these types of destruction rather than relying solely on a collapsing null shell with angular momentum to dissolve the BH horizon. One could first test the feasibility of BH destruction under a chosen scenario and then examine the observability using any feasible conditions, as not all BH models can be destroyed in this way~\cite{Wald:1974hkz,Barausse:2010ka,Sorce:2017dst, Cardoso:2015xtj, Jiang:2020mws, Yang:2022yvq}. This approach would allow for more accurate predictions of observables, potentially guiding future observations. Moreover, as discussed in Sec.~\ref{sec. image transition}, it would be valuable to implement a ray tracing method with finite resolution to account for the sudden burst of light produced at the moment the horizon disappears and to investigate its impact on actual observations.

We also assumed a thin accretion disk model with a GLM emission profile, along with a detection algorithm designed to sum the disk's intensity onto the image frame. While the GLM profile is widely used in studies of BH and ultracompact object appearances, it may not be fully appropriate for the superspinar scenario, which involves extreme spin conditions and could lead to distinct disk dynamics. Furthermore, employing a full radiative transfer treatment and a thick accretion disk model could result in a smoother image transition and potentially slightly lower observed intensity due to absorption and scattering within the disk material.

Nevertheless, we believe that these limitations do not significantly impact the validity of our main results.

\section*{Acknowledgements}

We thank Byon Jayawiguna for the enlightening discussions. We also thank the anonymous referee for their valuable insights and constructive feedback on this manuscript. HSR is funded by Hibah Matching-Fund UI-UNTAN No.~PKS-002/UN2.F3.D/PPM.00.02/2024. Hasanuddin is supported by Dana DIPA Universitas Tanjungpura No. SP DIPA-023.17.2.677517/2024.

\begin{appendix}

\section{Full expression of the Kretschmann scalar}

With the spacetime, mass function, and spin function given by Eqs.~\eqref{eq. null coord},~\eqref{eq. mass ghosh}, and~\eqref{eq. spin func}, respectively, the Kretschmann scalar $K\equiv R^{\mu\nu\alpha\beta}R_{\mu\nu\alpha\beta}$ is expressed as
\label{appendix. kretschmann}
\begin{align}
    K=&\frac{M^2e^{-\frac{2 k}{r}}}{4r^6 \Sigma^6}
    \left\{16A_1 r^8  + A_2M^{-2}e^{\frac{2 k}{r}} r^6- 16A_3 M^{-1} e^{k/r} r^3\right.\notag\\
    &+ a(u)^2 \cos ^2(\theta )\left[-2880 r^{10} + 1920 k r^9 + 384 k^2 r^8 - 384 k^3 r^7 + 64 k^4 r^6\right]\notag\\
    &+a(u)^4 \cos ^4(\theta )\left[2880  r^8 + 1920 k r^7 -384 k^2 r^6 + 96 k^4 r^4 - 384 k^3 r^5\right]\notag\\
    &+a(u)^6 \cos ^6(\theta )\left[-192 r^6 - 384 k r^5 -384 k^2 r^4 - 128 k^3 r^3 + 64 k^4 r^2\right]\notag\\
    &+16 k^4\left.a(u)^8 \cos ^8(\theta ) \right\},
\end{align}
where
\begin{align*}
    A_1=&k^4-8 k^3r+24 k^2r^2-24 kr^3+12 r^4,\\
    A_2=&2 B_1 a'(u)^2-\left[4B_2 r a(u)  a'(u)^3+8B_3r  a'(u)a''(u)\right]\sin^2(\theta)\\
    &+ \left[B_4 a'(u)^4 + 8B_5 a(u)^2  a''(u)^2\right]\sin^4(\theta),\\
    A_3=&B_6 r^7a(u)+B_7 r^5 a(u)^3\cos ^2( \theta )+B_8 r^3a(u)^5 \cos ^4(\theta )\\
    &+B_9 ra(u)^7 \cos ^6(\theta )+B_{10}a'(u)^2\sin^2 (\theta),
\end{align*}
and
\begin{align*}
    B_1 = & (-16 \cos (2 \theta )+\cos (4 \theta )-9)a(u)^8\cos ^6(\theta ) +2(-8 \cos (2 \theta )+\cos (4 \theta )+39) r^2 a(u)^6 \cos ^4(\theta )  \\
    &+(100 \cos (2 \theta )-9 \cos (4 \theta )+85)r^4 a(u)^4 \cos ^2(\theta )  +4(6 \cos (2 \theta )+\cos (4 \theta )+9) r^6 a(u)^2 \\
    &-8(\cos (2 \theta )+2) r^8  + a''(u)\sin^4(\theta)\left[-4 a(u)^9 \cos ^8(\theta )  +24 r^2 a(u)^7 \cos ^6(\theta ) \right.\\
    &+\left.32 r^4 a(u)^5 \cos^4(\theta)  +8 r^6 a(u)^3 \cos ^2(\theta )   +4 r^8  a(u)\right],\\
    B_2= & (5 \cos (2 \theta )+7) r^6 + 9(3 \cos (2 \theta )+1) r^4a(u)^2\cos^2(\theta)  \\
    &+(21-17 \cos (2 \theta ))r^2a(u)^4 \cos ^4(\theta )  -(7 \cos (2 \theta )+13)a(u)^6 \cos ^6(\theta ), \\
    B_3=& 2 r^8+(7 \cos (2 \theta )+9)r^6a(u)^2  + (21\cos(2\theta)+23)r^4a(u)^4 \cos^2(\theta)\\
    &+ (17 \cos (2 \theta )+31) r^2 a(u)^6 \cos ^4(\theta ) +2(\cos (2 \theta )+8) a(u)^8 \cos ^6(\theta ),\\
    B_4 = &3 r^8+4  r^6a(u)^2 \cos ^2(\theta )+74 r^4a(u)^4 \cos ^4(\theta ) +20 r^2a(u)^6 \cos^6(\theta ) +11 a(u)^8 \cos^8(\theta ),\\
    B_5=&r^8+4 r^6a(u)^2 \cos ^2(\theta ) +16r^4 a(u)^4 \cos ^4(\theta )+8 r^2a(u)^6 \cos ^6(\theta ) +3 a(u)^8 \cos ^8(\theta ),\\
    B_6 = & 2(k-r) r^2  a''(u) \sin ^2(\theta )+a'(u)\left[-k^2-6 kr+28 r^2+(k^2 - 14 k r + 24 r^2) \cos (2 \theta )\right],\\
    B_7=& 6(k+r) r^2 a''(u) \sin ^2( \theta ) + a'(u) \left[-3 k^2-16 kr-4 r^2+\left(3 k^2-12 kr-40 r^2\right) \cos (2 \theta )\right], \\
    B_8=&2 (3 k+r) r^2  a''(u) \sin ^2(\theta )+a'(u)\left[-3 k^2+2 kr-84 r^2+(3 k+2 r)k \cos (2 \theta )\right], \\
    B_9=&2(k-3 r) r^2  a''(u) \sin ^2(\theta )+a'(u)\left[ -2k^2 \sin^2(\theta)+12 kr+12 r^2\right],\\
    B_{10}=&2 (k^2-6 k r+8 r^2) r^8+\left(2 k^2-5 r k-24 r^2\right) r^6 a(u)^2 \cos ^2(\theta ) \\
    & 
    +\left(3 k^2-r k+28 r^2\right)r^4 a(u)^4 \cos ^4(\theta ) + \left(2 k^2+r k-8 r^2\right)r^2 a(u)^6 \cos ^6(\theta ) \\
    & + 2 k^2 a(u)^8 \cos ^8(\theta ).
\end{align*}
It should be noted that the prime ($'$) on $a(u)$ denotes differentiation with respect to $u$. For a stationary BH, \textit{i.e.} $a'(u)=a''(u)=0$, the terms $A_2$ and $A_3$ vanish. For our chosen form of $a(u)$ in Eq.~\eqref{eq. spin func}, the first and second derivatives are given by
\begin{align}
    a'(u)=&\frac{\delta}{2\sigma_s}\left[1+\cosh{\left(-\frac{u-u_c}{\sigma_s}\right)}\right]^{-1},\notag\\
    a''(u)=&\frac{2\delta}{\sigma_s^2} \operatorname{csch}^3\left(-\frac{u-u_c}{\sigma_s}\right)\sinh^4\left(-\frac{u-u_c}{\sigma_s}\right).
\end{align}
It can be seen that the presence of $a''(u)^2$ and $a'(u)^4$ in $A_2$ implies that the curvature scalar scales as $\sigma_s^{-4}$.

On the other hand, solving the field equations $G_{\mu\nu} = 8\pi T_{\mu\nu}$ with the metric in Eq.~\eqref{eq. null coord} yields all components of the energy-momentum tensor, with the $T_{uu}$ component given by
\begin{equation}
    T_{uu}=-\frac{4C_1+4a''(u)C_2+4a(u) a'(u)C_3+a'(u)^2C_4}{8\Sigma^{4}},
\end{equation}
where
\begin{align*}
    C_1=&2 r a(u)^2 \sin ^2(\theta ) m''(r) \Sigma+m'(r) \left[a(u)^4 \sin ^2(2 \theta )-4 r^2 \Xi\right],\\
    C_2=&D_1ra(u)^3 \cos ^2(\theta )+r^4 a(u) (\cos (2 \theta )+3)+4 a(u)^5 \cos ^4(\theta ),\\
    C_3=&D_2r^4-2 D_3r^2 a(u)^2 \cos ^2(\theta )+D_4a(u)^4 \cos ^4(\theta ),\\
    C_4=&D_5r^5+D_6r^3 a(u)^2 \cos ^2(\theta )+D_7r a(u)^4 \cos ^4(\theta )+10 a(u)^6 \sin ^2(\theta ) \cos ^6(\theta ),
\end{align*}
and
\begin{align*}
D_1=&r (\cos (2 \theta )+7)-8 \sin ^2(\theta ) m(r),\\
D_2=&\cos (2 \theta ) \left[12 m(r)-r \left(2 m'(r)+3\right)\right]+r \left[6 m'(r)-1\right],\\
D_3=&r \left[\cos (2 \theta )-4 m'(r)+3\right]+(5 \cos (2 \theta )-13) m(r),\\
D_4=&4 \cos ^2(\theta ) \left[r m'(r)+m(r)\right]+r (\cos (2 \theta )-5),\\
D_5=&r (3 \cos (2 \theta )+13)+12 \sin ^2(\theta ) m(r),\\
D_6=&r (39-7 \cos (2 \theta ))-88 \sin ^2(\theta ) m(r),\\
D_7=&r (31-15 \cos (2 \theta ))-4 \sin ^2(\theta ) m(r).
\end{align*}
For a stationary BH, we only left with $T_{uu}=-C_1/2\Sigma^4$.

\section{Spin transition rate ($\sigma_s$)}
\label{appendix. transition rate}

In Eq.~\ref{eq. spin func}, we introduce a quantity $\sigma_s$, which characterizes the inverse of the spin transition rate from $a_c \to a_c + \delta$. For an instantaneous destruction, one would have $\sigma_s \to 0$, which poses numerical challenges due to the appearance of a Dirac delta function in the derivative. Here we investigate how the choice of $\sigma_s$ would influences the null geodesic near the moment of destruction and its effects to the image transition, particularly in $\delta/M=5\times10^{-3}$, for four representative $\sigma_s$ values: $\sigma_s \in \{10^{-1}, 10^{-2}, 10^{-3}, 10^{-4}\}$.

\begin{figure}[htbp!]
    \centering
\includegraphics[width=1\linewidth]{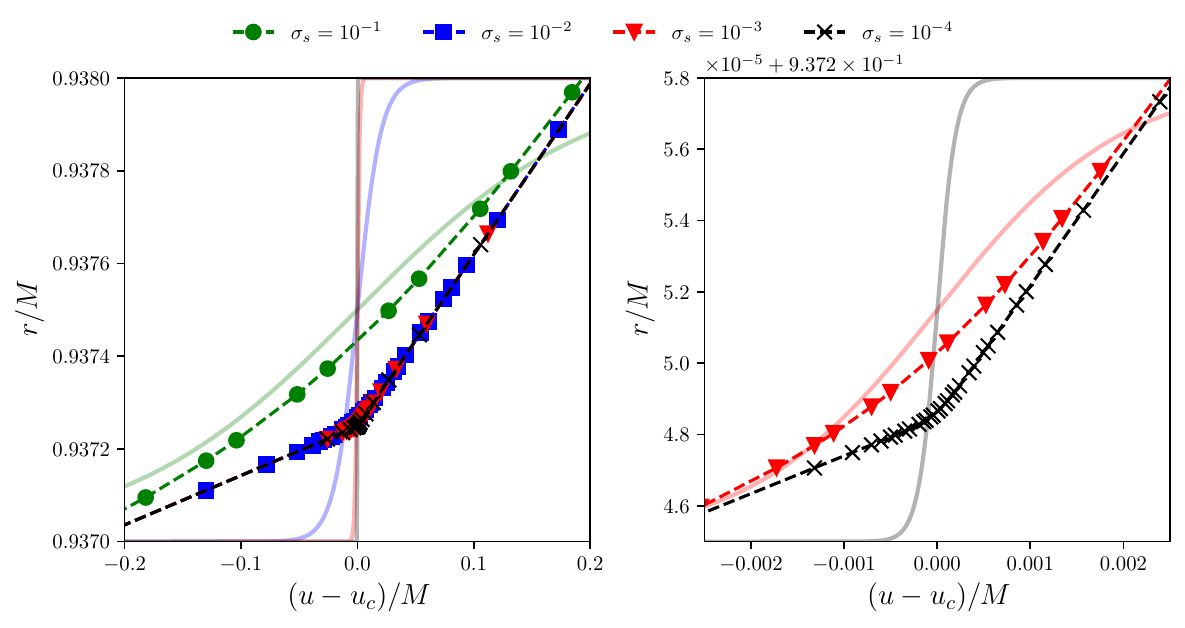}
    \caption{Radial null trajectories near $u_c$ for several values of $\sigma_s$ considered in Sec.~\ref{sec. image transition} for $\delta/M=5\times10^{-3}$ emitted at $u=-50M$ from $r=10M$. The faint solid lines represent the $\Theta(u-u_c)$ for each corresponding $\sigma_s$ normalized to the vertical axis of the plot.}
    \label{fig. varying sigma}
\end{figure}

Let us take a closer look at the radial null geodesics near $u=u_c$, shown in Fig.~\ref{fig. varying sigma}. One can see that the null geodesics converge for small $\sigma_s$, and variations in $\sigma_s$ produce identical trajectories for $|u-u_c|\gg 0$. This behavior is also confirmed for non-radial trajectories. Close to the collapsing null shell, these photons are ``pushed" outward, resulting in a steeper $dr/du$. Moreover, our geodesic integration method successfully accounts for the narrow derivative even at small $\sigma_s$, as indicated by the tighter integration steps marked along the curves, which sufficiently resolve the region near the steep $\partial_u a(u)$.

Accordingly, since all values of $\sigma_s$ yield the same geodesics for $|u-u_c|\gg 0$, we confirm that they produce essentially identical image and intensity transitions of the BH destruction. Therefore, we conclude that our sigmoid function approach for the dynamical spin in Eq.~\eqref{eq. spin func} provides an excellent approximation, as the resulting transitions are largely independent of the choice of $\sigma_s$.

\section{Imaging the destruction with \textit{tricks} (our initial approach)}

\begin{figure}[htbp!]
    \centering
\includegraphics[width=0.8\linewidth,trim={0 5cm 0 0}]{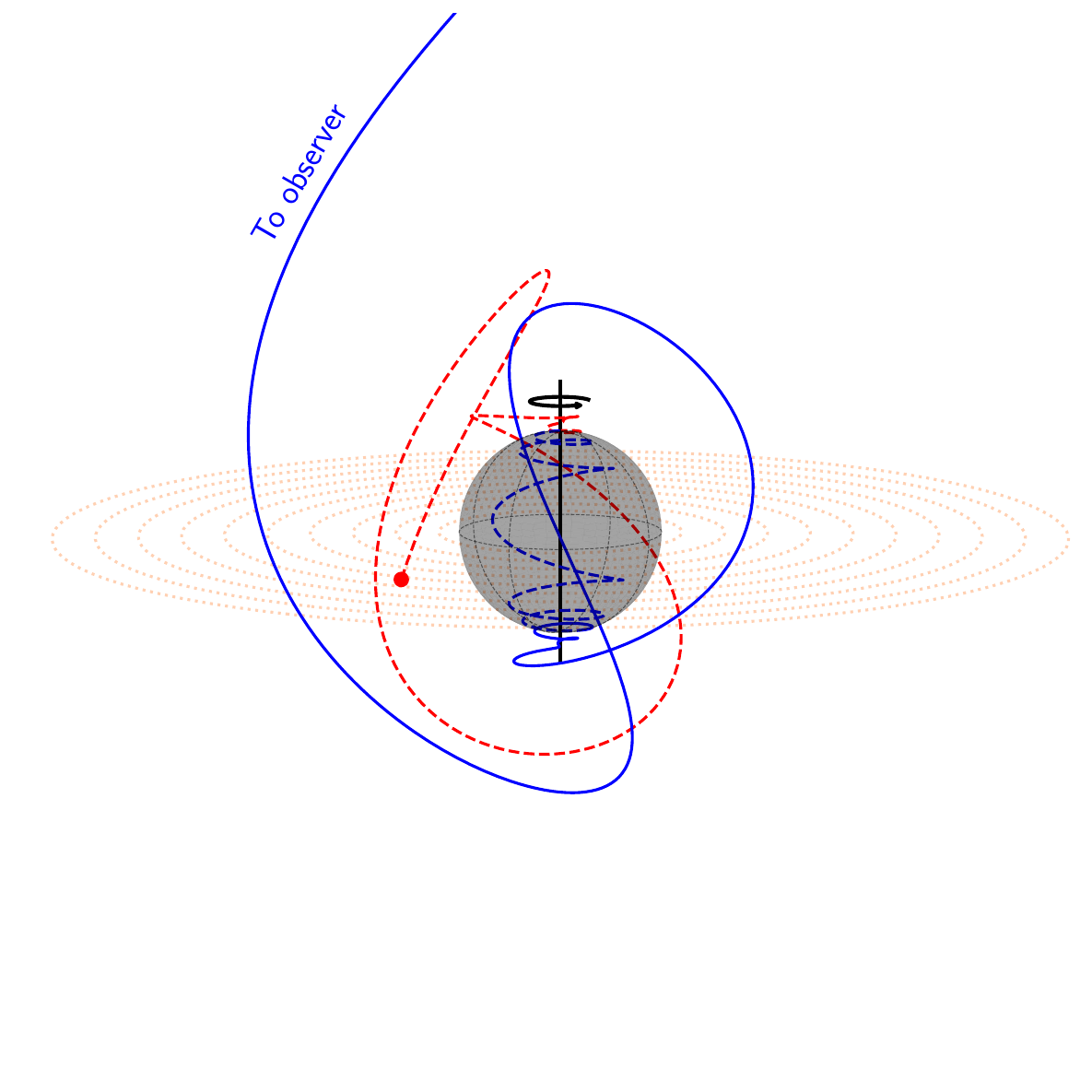}
    \caption{The photon trajectory and its recorded travel time mechanism, depicting their journey to the observer. A photon is emitted from the accretion disk, originating from the red solid circle, and travels through the ``imaginary" horizon (represented by the greyed spherical region) before reaching the observer. Our algorithm records only the travel time associated with the blue lines: the dashed line represents the photon’s trajectory inside the imaginary horizon (interior of the superspinar), while the solid line indicates its path after exiting this region.}
    \label{fig. geodesic time}
\end{figure}

In the first approach, we seek a fast and efficient method to compute the transition image. We employed a trick in the ray tracing: instead of solving the null geodesics in a dynamical spacetime, we use the stationary spacetime corresponding to the final state of the object (a superspinar with $a = a_c + \delta$), while imposing a \textit{stopping condition} at the extremal horizon radius for the initial image at $t=0$, despite the absence of an actual event horizon. This provides a practical framework for image generation that efficiently captures the formation of secondary images within the previously shadowed region. Figure~\ref{fig. geodesic time} illustrates how the travel time of each ray is recorded. The entire blue and red paths represent actual photon trajectories, obtained by solving the corresponding null geodesic equations, Eq.~\eqref{eq. dH x}.

\begin{figure}[htbp!]
    \centering
\includegraphics[width=0.7\linewidth]{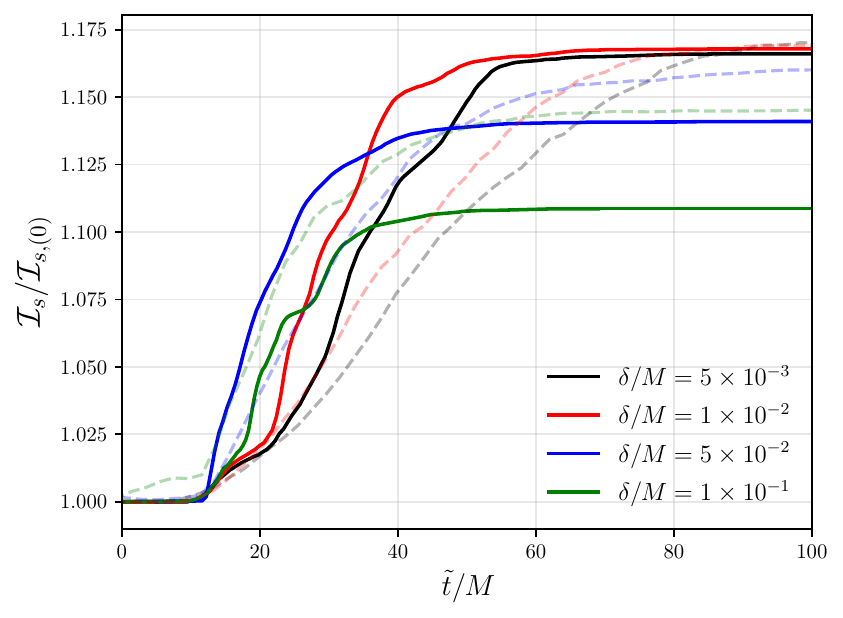}
    \caption{Comparing the results of the integrated intensity $\mathcal{I}_s$ evolution over shifted time $\tilde{t}$ with tricks (solid lines) and our current results (dashed transparent lines) for values of $\delta$ used in the main discussion.}
    \label{fig. compare tricks}
\end{figure}

Initially, we found that this trick appeared to be a convincing approach to image the BH’s destruction for small $\delta$. The image transition could be obtained efficiently in a single computation, and the intensity evolution seemed to agree with our initial expectation (see Fig.~\ref{fig. compare tricks}). However, we later realized that this trick is plagued by an unacceptable inconsistency.

As the instantaneous spin transition $a \to a + \delta$ occurs at a particular coordinate time $t$ rather than the null coordinate $u$, it implies that the spacetime is changing non-locally. This trick also fails to account for the photon geodesic behavior near\textemdash or even inside\textemdash the BH at the moment of its destruction, which may be crucial for accurately capturing the intensity evolution. At relatively large $\delta$, this approach leads to entirely different null geodesics around the object due to the significant spin deviation from the BH limit, resulting in inconsistencies in both the images and the intensity evolution. Even so, this method can reproduce a similar effect by varying $\delta$, where smaller $\delta$ leads to a longer transition time, although it produces a transition approximately $\sim 1.75$ times faster than our current results.

\end{appendix}

\end{document}